\documentclass[11pt]{article}
\usepackage{jheppub}

\usepackage{epsfig}
\usepackage{amssymb}
\usepackage{amsfonts}
\usepackage{amsbsy}
\usepackage[all,v2,cmtip,2cell]{xy}
\usepackage{amsmath}
\usepackage{dsfont}
\usepackage{bbm}
\usepackage{upgreek}
\usepackage{amssymb,amscd}
\usepackage{graphicx}
\usepackage{mathrsfs}
\usepackage{amsmath,amsthm}
\usepackage{slashed}
\usepackage{dsfont}
\usepackage[utf8]{inputenc}
\makeatletter
\DeclareFontFamily{OMX}{MnSymbolE}{}
\DeclareSymbolFont{MnLargeSymbols}{OMX}{MnSymbolE}{m}{n}
\SetSymbolFont{MnLargeSymbols}{bold}{OMX}{MnSymbolE}{b}{n}
\DeclareFontShape{OMX}{MnSymbolE}{m}{n}{
    <-6>  MnSymbolE5
   <6-7>  MnSymbolE6
   <7-8>  MnSymbolE7
   <8-9>  MnSymbolE8
   <9-10> MnSymbolE9
  <10-12> MnSymbolE10
  <12->   MnSymbolE12
}{}
\DeclareFontShape{OMX}{MnSymbolE}{b}{n}{
    <-6>  MnSymbolE-Bold5
   <6-7>  MnSymbolE-Bold6
   <7-8>  MnSymbolE-Bold7
   <8-9>  MnSymbolE-Bold8
   <9-10> MnSymbolE-Bold9
  <10-12> MnSymbolE-Bold10
  <12->   MnSymbolE-Bold12
}{}

\let\llangle\@undefined
\let\rrangle\@undefined
\DeclareMathDelimiter{\llangle}{\mathopen}%
                     {MnLargeSymbols}{'164}{MnLargeSymbols}{'164}
\DeclareMathDelimiter{\rrangle}{\mathclose}%
                     {MnLargeSymbols}{'171}{MnLargeSymbols}{'171}
\makeatother

\def\be{ \begin{equation} }
\def\ee{ \end{equation}}

\newcommand{\eq}[1]{\begin{align}\begin{split}#1\end{split}\end{align}}

\def\tildeB{{\widetilde{B}}}
\def\tildeF{{\widetilde{F}}}
\def\tildeS{{\widetilde{S}}}
\def\tildeW{{\widetilde{W}}}
\def\tildeM{{\widetilde{M}}}

\def\exp{{\rm exp}}

\def\log{{\rm log}}

\def\half{\frac{1}{2}}

\def\itwopi{\frac{i}{2\pi}}

\def\one{{\hbox{ 1\kern-.8mm l}}}

\def\CA{{\cal A}}

\def\CB{{\cal B}}
\def\CC {{\cal C}}
\def\CD {{\cal D}}

\def\CG {{\cal G}}

\def\CI {{\cal I}}

\def\CK {{\cal K}}
\def\CL {{\cal L}}

\def\CN {{\cal N}}

\def\CO {{\cal O}}
\def\CP {{\cal P}}

\def\CW {{\cal W}}
\def\CX {{\cal X}}
\def\CO {{\cal O}}

\def\CG {{\cal G}}

\def\CI {{{\cal I}}}
\def\CB {{\cal B}}

\def\CS {{\cal S}}
\def\CT {{\cal T}}

\def\CX {{\cal X}}

\def\IC{\mathbb{C}}

\def\IG{\mathbb{G}}

\def\ICP{\mathbb{CP}}
\def\IQ{\mathbb{Q}}

\def\IZ{{\mathbb{Z}}}

\def\fB{\mathfrak{B}}

\def\rmk#1{\bigskip\noindent{\bf Remark} }
\def\cnj#1{\bigskip\noindent{\bf Conjecture:} }

\def\tildeG{{\widetilde{G}}}

\def\tildeA{{\widetilde{A}}}

\DeclareMathAlphabet{\mathpzc}{OT1}{pzc}{m}{it}

\def\Tr{ \, \textrm{Tr} \, }

\def\tildeb{{\tilde{b}}}
\def\tildec{{\tilde{c}}}

\title{Generalized Families of QFTs}

\author[1,2]{T.~Daniel Brennan}
\author[1]{and Kenneth Intriligator}

\affiliation[1]{Department of Physics, University of California San Diego,\\
 \textit{9500 Gilman Drive, La Jolla CA 92093-0319, USA}}
 \affiliation[2]{School of Mathematics, University of Birmingham\\
 \textit{Watson Building, Edgbaston, Birmingham B15 2TT, UK}
 }

\abstract{
RG flows and IR phases of QFTs can be constrained by generalized symmetries and their anomalies.  Broken symmetries act on the space of coupling constants of families of theories, and can also have IR-constraining family anomalies.  
We generalize family anomaly considerations to cases of broken generalized/categorical symmetries, including higher-group and non-invertible symmetries.  We consider the anomaly inflow and SymTFTs of such generalized families of QFTs, and their implications for RG flows and constraints on the IR phases.  As examples, we apply family anomalies to study the IR phases of $4d$ QCD-like theories deformed by irrelevant, multi-fermion interactions. 
}

\begin{document}

\maketitle

\section{Introduction}
Global symmetries, and their anomalies, can powerfully constrain the renormalization group  and IR phases of quantum field theories.    Generalized global symmetries~\cite{Gaiotto:2014kfa} include all topological operators, which can act on extended operators (line-operators, etc.) in QFTs.   Generalized anomalies arise as a tension between the generally projective action of symmetries on Hilbert space vs the linearity of quantization~\cite{Freed:2023snr}.  
There are also possible anomalies in the space of coupling constants, or \emph{family anomalies}~\cite{Cordova:2019jnf, Cordova:2019uob}, which can extend symmetry-based constraints to theories deformed by interactions that explicitly break the symmetries. 

We will  discuss generalized symmetry variants of family anomalies, and their implications for IR phases.  Much as with the original 't Hooft anomalies~\cite{tHooft:1979rat,Frishman:1980dq,Coleman:1982yg}, such generalized symmetry anomalies can provide new constraints on dynamics and RG flows and exclude IR phases, or to infer phenomena including symmetry fractionalization etc; see e.g.  \cite{Lieb:1961fr,Wang:2014lca,Wang:2016gqj,Wang:2017loc,Kobayashi:2018yuk,Wan:2018djl,Wan:2019oyr,Cordova:2019bsd,Cordova:2019jqi,Apte:2022xtu,Brennan:2023kpo,Brennan:2023ynm,Brennan:2024tlw,Hsin:2020cgg,Zhang:2023wlu,Cordova:2023bja,Antinucci:2023ezl,Jackiw:1975fn,Barkeshli:2014cna,Delmastro:2022pfo,Lee:2021crt,Brennan:2022tyl,Brennan:2025acl,Wang:2013zja,Thorngren:2014pza,Kravec:2014aza,Zou:2017ppq,Wang:2018qoy,Hsin:2019fhf,Kan:2024fuu,Damia:2023ses,Antinucci:2023ezl} and references therein for examples. 

Generalized symmetries go beyond group theory, to category theory.  See e.g.~\cite{McGreevy:2022oyu,Cordova:2022ruw,Schafer-Nameki:2023jdn,Shao:2023gho,Bhardwaj:2023kri,Brennan:2023mmt,Costa:2024wks} for reviews. The symmetry category and their anomalies can be nicely encapsulated by a TQFT in one-higher dimension:  i.e. the anomaly SPT or 
the ``SymTFT'' \cite{Gaiotto:2020iye,Freed:2022qnc,Kaidi:2021xfk,Kaidi:2022cpf,Kaidi:2023maf,Kong:2020cie,Apruzzi:2021nmk,Bhardwaj:2023bbf,Apruzzi:2023uma,Bhardwaj:2023wzd,Bhardwaj:2023ayw,Bartsch:2023pzl,Bartsch:2023wvv,vanBeest:2022fss,Sun:2023xxv,Zhang:2023wlu,Cordova:2023bja,Antinucci:2023ezl,Brennan:2024fgj,Antinucci:2024ltv,Antinucci:2024zjp,Bonetti:2024cjk,DelZotto:2024tae,Apruzzi:2024htg,Baume:2023kkf,Chang:2018iay,Choi:2021kmx,Roumpedakis:2022aik,Bhardwaj:2022yxj,Choi:2022zal}. The SymTFT describes the symmetry of a $d$-dimensional QFT${}_d$ on a spacetime $M_d$ as a TQFT on $[0,1]_t\times M_d$, where the QFT${}_d$ resides at $t=0$.  In this framework a topological, gapped boundary condition at $t=1$ controls how the symmetry category acts on the Hilbert space.

We will consider continuous families of $d$-dimensional QFTs $\CT_\theta$ (indexed by continuous $\theta\in \CC_\theta$) with a fixed, preserved $\CG$ global symmetry, and an additional symmetry $\CK$ that is explicitly broken by some of the interactions in the family of couplings. Though the couplings explicitly break $\CK$, one can still get symmetry-based selection rules if we assign appropriate $\CK$-transformations to the couplings.  Regarding the couplings as the expectation values of fictitious {\it spurion} fields, the explicit breaking can be rephrased as spontaneous symmetry breaking from the vev of the charged spurion. If we restrict attention to families with a fixed, preserved global symmetry, the parameter space can have non-trivial topology such as non-contractible loops.   We consider $\CT_\theta$ parametrized by 
a continuous, dimensionless parameter $\theta$ that shifts under broken symmetry transformations.    In terms of  spurions,  $\theta$ is the corresponding, would-be Nambu-Goldstone boson for $\CK$ when $\CK$ is continuous.  E.g. $\theta$ could be the coefficient of $\Tr F\tilde F$ in $4d$ gauge theories, or the phase of a complex parameter such as a mass term.  We assume that $\CK$ is compact, which leads to periodic identifications of $\theta$, e.g. $\CT_\theta\cong \CT_{\theta+2\pi}$. Such periodicities in the space of coupling constants can have associated anomalies, as in~\cite{Cordova:2019jnf, Cordova:2019uob}, when the partition functions differ by a phase:
\eq{
Z_{\theta}[A]=e^{i \int\omega_d(A)}\times Z_{{\theta+2\pi}}[A]~.
}
The possible {\it family anomaly} $\omega_d$ is an integer quantized characteristic class for the background gauge fields $A$ for $\CG$. The quantization condition is such that $\exp\left\{ i k \int_{M_d} \omega_d(A)\right\}$ is only invariant under $\CG$ background gauge transformations if $k\in \IZ$. 
When $\omega_d\neq 0$,  the partition function is a section of a non-trivial complex line bundle over the parameter space $\CC_\theta$ reduced by the broken symmetry generators. The family anomaly  can be encoded by a $(d+1)$-dimensional SPT phase/anomaly theory with path integrand $e^{i{\cal A}}$ with action
\eq{
\CA = \int \frac{d\theta}{2\pi}\cup \omega_d~,
}
where $\theta$ is lifted to a $\CC_\theta\cong U(1)$-valued scalar field.  For example, $SU(N)$ Yang Mills in $d=4$ has an anomaly in $\theta _{YM}\sim \theta _{YM}+2\pi$ in the presence of a background field $B^{(2)}$ for the $\IZ _N^{(1)}$ global symmetry~\cite{Cordova:2019jnf, Cordova:2019uob}
\eq{\CA \subset  \frac{N-1}{2N}\int d\theta _{YM} \cup {\cal P}(B)~.}

As with other 't Hooft anomalies, the family anomaly SPT is scale invariant and thus the family anomaly must be matched along RG flows.  Non-zero family anomalies imply that the family flows to a non-trivial family of theories in the IR, e.g. that there exists some $\theta_0\in \CC_\theta$ where the IR theory is not in a trivially gapped phase~\cite{Cordova:2019jnf,Cordova:2019uob}. This can imply a phase transition, with the Hilbert space not uniform over $\CC_\theta$. See e.g.~\cite{Debray:2023ior,Lichtman:2020nuw,Brennan:2024tlw,Choi:2022odr,Hsin:2020cgg} for additional examples and discussion.   
Continuous families of theories can also arise by compactifying theories with continuous symmetries, with $\theta$ the holonomy of continuous background gauge fields.   The family anomalies can thus descend from the anomalies of the continuous symmetry that generates the twisted compactification; see e.g. \cite{Kobayashi:2023ajk,Nardoni:2024sos} for recent discussion.

We will here consider extending these notions and applications to the cases with categorical symmetries acting on families of QFTs, as can occur when  categorical symmetry is explicitly broken by a family of coupling constant deformations.  Such   
\emph{generalized families of theories} can be thought of as the structure inherited by explicitly broken higher group or non-invertible chiral symmetries. For generalized families of theories, the symmetry action on the coupling constant parameter leads to a generalized family relation analogous to the periodicity of $\theta$ from standard, continuous families of theories. Such categorical families of theories are the categorical extension of the notion of ``$(-1)$-form global symmetries'' which was discussed in e.g. \cite{Cordova:2019jnf,Cordova:2019uob,Heidenreich:2020pkc,Albertini:2020mdx,McNamara:2020uza,Sharpe:2022ene,Vandermeulen:2022edk,Vandermeulen:2023smx,Aloni:2024jpb,Yu:2024jtk,Santilli:2024dyz,Najjar:2024vmm,Seiberg:2024yig,Lin:2025oml,Robbins:2025apg,Seiberg:2025bqy}.   We will here only consider cases where the family structure includes a broken $U(1)\cong S^1$ (or a subgroup), with associated parameter $\theta \sim \theta +2\pi$.  Parameter spaces with higher-dimensional, non-trivial topologies can lead to other interesting effects that have been especially considered in condensed matter theory contexts  see e.g.~\cite{Hsin:2020cgg, Manjunath:2024rxe,Choi:2025ebk,Kapustin:2020eby,Debray:2023ior,Wen:2021gwc,Hsin:2022iug} and references therein for recent discussion and examples.

A classic example of a non-invertible symmetry is the $\IZ_ 2$ Kramers-Wannier (KW) duality of the $(1+1)d$ Ising model at its critical point,  see e.g.~\cite{Shao:2023gho} and references therein.  The KW symmetry is explicitly broken when the theory is deformed away from the critical point, and then the broken symmetry acts on the family of non-critical theories.  This broken symmetry action is indeed the original Kramers-Wannier duality map away from the critical point, between the high-temperature and low-temperature phases, or  $m\to -m$ in the fermion description.  KW duality is an example of an explicitly broken, non-invertible family symmetry.

As in the case of general symmetries, family relations can be accidental (i.e. an isomorphism of low energy effective theories with different coupling constants but not of the full UV theory).   
The nested structure of the categorical symmetry implies a necessary hierarchy of energy scales along the RG flow for accidental categorical symmetries.  This can lead to universal inequalities on coupling constants and constraints on UV completions, see e.g. \cite{Cordova:2018cvg,Brennan:2020ehu,Choi:2022fgx,Brennan:2023kpw}. 
Similarly, emergent categorical family structure may also require the existence of a hierarchy of energy scales  which can be used to give universal constraints on UV completions.

We will discuss the anomalies of generalized/categorical families and their constraints on RG flows and the IR theory. We will find that the constraints on RG flows will be similar to those of higher group and non-invertible symmetries:
\begin{itemize}
\item Higher family anomalies, when non-trivial, imply that the IR theory either spontaneously breaks the global symmetry or has a phase transition.
\item The anomalies of non-invertible family symmetries are trivialized if the anomaly coefficient satisfies an algebraic constraint similar to those derived in \cite{Apte:2022xtu,Cordova:2023bja,Damia:2023ses,Antinucci:2023ezl}.  If the trivializing condition is not satisfied, the non-invertible family anomaly {\it cannot} be cancelled by dressing the theory by an SPT counterterm -- i.e. it is a genuine anomaly.  
\end{itemize}

As an application of these concepts, we study examples with irrelevant, categorical symmetry breaking deformations in QCD-like gauge theories in $4d$.  The categorical symmetry is then accidental in the deep IR, and along the RG flow to the IR it can lead to a categorical family structure that constrains the RG flow's phase structure to 
the IR family of theories.    

\vspace{0.25cm}
The outline of this paper is as follows. We review generalized symmetries in Section \ref{sec:review}, and the notion of generalized families of theories in Section \ref{sec:genfam}.  We discuss the anomalies of categorical families in Section \ref{sec:anomalies}. In Section \ref{sec:nequals2} we illustrate generalized families of theories and their anomalies in the context of an extended example: $4d$ $\CN=2$ $SU(2)$ and $SO(3)$ SYM.  In Section \ref{sec:dangerouslyirrelevant} we further illustrate generalized families and related anomalies in several examples of QCD-like theories with various (irrelevant) deformations.  In Section \ref{app:NonInvertible}, we discuss when the higher family anomaly trivializes, similar to the constraints found in \cite{Apte:2022xtu,Cordova:2023bja,Damia:2023ses,Antinucci:2023ezl}. 

\section{Review of Generalized Global Symmetries}
\label{sec:review}

Any topological operator in a $d$-dimensional QFT corresponds to a generalized global symmetry~\cite{Gaiotto:2014kfa}, which can generally act on operators of dimension greater than or equal to the symmetry topological operator's co-dimension minus one.  A $p$-form global symmetry $G^{(p)}$ assigns charges to $p$-dimensional operators and is generated by the surrounding co-dimension $(p+1)$ topological  symmetry defect operators $U_g$.  They are labeled by elements $g\in G^{(p)}$ such that their fusion satisfies the group product law: $U_{g_1}\cdot U_{g_2}=U_{g_1g_2}$. The topological invariance of the symmetry defect operators implies that $p>0$ higher form symmetries $G^{(p)}$ must be abelian.  The  $p$-form global symmetry can be sourced by a background gauge field, which is locally a degree-$(p+1)$ differential form $A_{p+1}$, where $A_{p+1}$ transforms under the gauge transformations 
$\delta A_{p+1}=d\lambda_p$ where $\lambda_p$ is a $G^{(p)}$-valued $p$-form which is the gauge transformation parameter.

\subsection{Higher Group Global Symmetries}
\label{sec:highergroup}

Higher group global symmetries are an extension of higher form symmetries. A higher group of degree $n$ (or $n$-group) $\IG_n$ is composed of $p$-form global symmetries where $p$ runs from 0 to $n-1$. The higher form components can mix into each other, as reflected in non-trivial braiding/fusion of the corresponding topological operators. The structure of the higher group can be captured by the gauge transformations of the background gauge fields of its constituent higher form global symmetries:
\eq{\label{generalHGtrans}
\delta A_{p+1}=d\lambda_p+\sum_{q\leq p}\lambda_q \wedge \omega_{p-q+1} (A,\lambda)~,
}
where $\omega_{k}(A,\lambda)$ is a $k$-form characteristic class that can depend on the background gauge fields $A$ and background transformation parameters for the $G^{(q)}$ for $q\leq k$. In other words, the $G^{(q)}$ transformations mix with the $G^{(p)}$ background gauge fields for $q<p$. 

An $n$-group $\IG_n$ has a natural nested sub-$n$-group structure
\eq{
\IG_n\supseteq \IG_n^{(1)}\supseteq \IG_n^{(2)}\supseteq...\supseteq \IG_n^{(n-1)}=G^{(n-1)}~,
}
where $\IG_n^{(p)}$ is the $n$-subgroup that contains only $q>p$-form components. As discussed in \cite{Cordova:2018cvg,Brennan:2020ehu}, the nested structure implies that if a given QFT has a $\IG_n$ global symmetry that becomes emergent along an RG flow, then its emergence must respect the nested higher group structure: 
\eq{
E_{\IG_n}\lesssim E_{\IG_n^{(1)}}\lesssim E_{\IG_n^{(2)}}\lesssim...\lesssim E_{G^{(n-1)}}~,
}
where $E_{\IG_n^{(p)}}$ is the energy scale below which the $\IG_n^{(p)}$ $n$-group global symmetry emerges. 

Simple, non-trivial examples of 2-group global symmetry arise in variants of 4d QED~\cite{Cordova:2018cvg}. Consider e.g. 4d QED with $N_f$ Dirac fermions of charge 1, i.e. $N_f$ Weyl fermions of charge $+1$ and $N_f$ Weyl fermions of charge $-1$.  The global symmetry includes a $U(1)_m^{(1)}$ magnetic 1-form global symmetry and $SU(N_f)_L\times SU(N_f)_R$ global symmetry.   There is also a  non-invertible global symmetry from subgroups of the ABJ-anomalous $U(1)_A$ symmetry~\cite{Choi:2022jqy,Cordova:2022ieu}, which we will review and further discuss later.  The $SU(N_f)_{L-R}$ global symmetry combines with $U(1)_m^{(1)}$ to form a 2-group~\cite{Cordova:2018cvg}.  The  
$U(1)_m^{(1)}$ magnetic 1-form global symmetry can be coupled to a background 2-form gauge field $B_2^{(m)}$ via a term 
\eq{
S\supset \itwopi \int B_2^{(m)}\wedge F_2~, 
}
where $F_2$ is the field strength of the dynamical $U(1)_g$ gauge field. The 2-group Postnikov class cancels the apparent mixed anomaly between $SU(N_f)_L$ (more generally, $SU(N_f)_{L-R}$) and the $U(1)_g$ gauge symmetry
\eq{ \kappa _{SU(N_f)_LU(1)_g}={\rm Tr}\ [SU(N_f)_L^2\times U(1)_g]=1~. 
}
This shows that  the path integral measure varies under $SU(N_f)_L$ background gauge transformations $\delta A_L=D\lambda _L$.   
This variation can then be canceled by assigning 2-group transformations of the background gauge field for $U(1)^{(1)}_m$
\eq{\label{2grouptransexample}
\delta B_2^{(m)}=d\Lambda_1+\frac{1}{4\pi}\Tr (\lambda _L \,dA_L)~,
}
deforming the classical symmetry to a 2-group global symmetry~\cite{Cordova:2018cvg}.

\subsection{Non-Invertible Chiral Symmetries}
Another generalization of the simple group-like global symmetry are non-invertible global symmetries. We will here only consider non-invertible symmetries that are similar to that in $4d$ massless QED, as discussed in~\cite{Choi:2022jqy,Cordova:2022ieu}. These are global symmetries that are naively broken by an ABJ anomaly and then recovered (at least for a possibly dense subgroup) by dressing the broken symmetry defect operator by a non-trivial TQFT that locally gauges a non-anomalous global symmetry of the full theory. The dressing by non-trivial TQFTs can be interpreted as implementing a  discrete gauging across the charge's world-volume; this cancels the ABJ anomaly, at the expense of making the fusion rules non-invertible. 
 
Following~\cite{Choi:2022jqy,Cordova:2022ieu}, consider the example of $4d$ QED with a single massless Dirac fermion. There is a classical $U(1)_5$ chiral symmetry whose associated current $j_5$ has an ABJ anomaly
\eq{U(1)_5:\qquad 
\psi\longmapsto e^{i \alpha \,\gamma _5}\psi~, \qquad \qquad d\ast j_5=\frac{F_2\wedge F_2}{4\pi ^2}~,
}
where $F_2$ is the field strength of the dynamical $U(1)_g$ gauge field. 
Because of the anomalous conservation equation, the naive symmetry defect operator $e^{ i \alpha \oint \ast j_5}$ is not topological. For $\alpha =2\pi q$ with $q\in \IQ/\IZ$, we can construct a topological operator 
\eq{
\CD_q(\Sigma)=e^{ 2\pi i q\oint_\Sigma \ast j_5}\, \CA^{N,p}\left[\Sigma;\frac{F_2}{2\pi}\right]~, 
}
where $q=p/N$ for ${\rm gcd}(p,N)=1$ and $\CA^{N,p}[\Sigma; B_2]$ is the minimal $3d$ $\IZ_N$ TQFT on $\Sigma$ of \cite{Hsin:2018vcg}. The TQFT $\CA^{N,p}\left[\Sigma;\frac{F_2}{2\pi}\right]$ gauges a $\IZ_N$ subgroup of the $U(1)^{(1)}_m$ 1-form magnetic symmetry on $\Sigma$ with a torsion term.  Another construction of the $\CA^{N,p}[\Sigma; B_2]$ TQFT is in terms of the half-space gauging construction of \cite{Choi:2021kmx,Choi:2022zal} which couples the QFT to a TQFT on one component of spacetime divided by $\Sigma$:
\eq{
\CA^{N,p}[\Sigma;B_2]~\Longrightarrow ~S_{TQFT}[B_2]=\int_{\sigma_+}\frac{iN}{2\pi} dc_1\wedge b_2+i b_2\wedge B_2+\frac{i kN}{4\pi}b_2\wedge b_2~, 
}
where $\oint B_2\in \IZ$, $c_1,b_2$ are $U(1)$-valued gauge fields of degree 1,2 respectively, $\sigma_+$ is the half-space with boundary $\partial \sigma_+=\Sigma$, and $kp=1$ mod$_{\gamma(N)N}$ where 
\eq{
\gamma(N)=\begin{cases}
1&N\text{ odd}\\
2&N\text{ even}
\end{cases}
}
This construction similarly gauges a subgroup of the $U(1)^{(1)}_m$ magnetic 1-form global symmetry of the full theory on one side of the operator $\CD_q(\Sigma)$. 

If we integrate out $b_2,c_1$, we find that this leads to an effective fractional $\theta$-angle: 
\eq{
S_{TQFT}\left[\frac{F_2}{2\pi}\right]~\longmapsto ~\frac{2\pi ip}{N}\int \frac{F_2\wedge F_2}{8\pi^2}~,
}
which cancels the term that is generated by the $U(1)_5$ rotation. The expense of this cancellation is that the topological operators no longer have an invertible fusion rule~\cite{Choi:2021kmx,Choi:2022zal}:
\eq{
\CD_{-1/N}(\Sigma)\cdot \CD_{1/N}(\Sigma)=\CC_0(\Sigma)~,
}
where $\CC_0(\Sigma)$ is a condensation defect which locally gauges $\IZ_N^{(1)}$ symmetry on $\Sigma$ \cite{Roumpedakis:2022aik}. 

Note that this construction  of non-invertible symmetries requires a higher form global symmetry $G$, that is gauged to cancel the anomalous conservation equation -- in the example above, it is $G=U(1)^{(1)}_m$. This implies that if the non-invertible symmetry is an accidental symmetry of an IR theory, then any RG flow which ends there must obey an energy hierarchy:
\eq{
E_{G}\gtrsim E_{\rm non\text{-}invertible}
}
where $E_G$ is the energy scale where the higher form symmetry $G$ that participates in the non-invertible symmetry emerges.

\subsection{SymTFT}

\begin{figure}  
    \centering
    \includegraphics[scale=1.2,clip,trim=0.5cm 23.5cm 12cm 0cm]{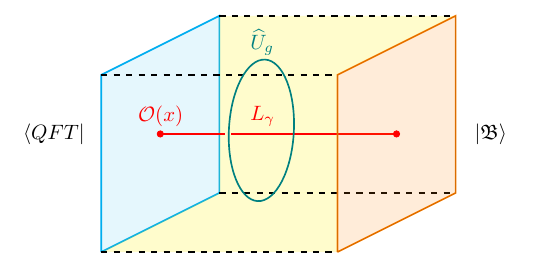}
    \caption{In this figure, we illustrate the idea of the SymTFT. Here, the left boundary (blue) is the QFT boundary (i.e. where the physical QFT exists) and the right boundary (orange) is the topological boundary which is labeled by a topological boundary condition $\mathfrak{B}$. In the SymTFT, these boundary conditions can also be written as states in the Hilbert space of the topological theory $\langle QFT|$ and $|\mathfrak{B}\rangle$ respectively. Here the symmetry operators $U_g$ of the QFT can be lifted directly to operators $\widehat{U}_g$ (teal) in the SymTFT whereas charged operators (red) $\mathcal{O}(x)$ are attached to topological operators $L_\gamma$ that run between the boundaries in the SymTFT.}
\label{fig:SymTFT}
\end{figure}

The abstract, full symmetry category of a $d$-dimensional QFT can be physically realized in terms of a $(d+1)$-dimensional TQFT called the ``symmetry TQFT'' or ``SymTFT'' for short. See~\cite{Freed:2022qnc,Schafer-Nameki:2023jdn,Shao:2023gho,Bhardwaj:2023kri,Freed:2022iao} for further details and discussion.  In brief, for a $d$-dimensional QFT on $M_d$, one couples  the theory to a TQFT on $Y_{d+1}=M_d\times [0,1]_t$.  The interval is parametrized by $t$, with the QFT at $t=0$ such that all symmetry operators $U_g$ in the QFT correspond to a topological operator $\widehat{U}_g$ in the TQFT that is sent to the boundary at $t=0$ and all charged operators $\CO(\Gamma)$ live at the end of 1-higher dimensional topological operators $L_\gamma$ where $\partial \gamma=\Gamma_{t=0}\cup \overline\Gamma_{t=1}$. See Figure \ref{fig:SymTFT}. The QFT can only be consistently coupled to the TQFT if the action of the symmetry operators on charged operators in the QFT is matched by the braiding and fusion of the purely topological operators in the SymTFT. This construction is a physical realization of the symmetry category of the given QFT.

The SymTFT construction requires a choice of (topological) boundary condition $\mathfrak{B}$ at $t=1$. This quiche~\cite{Freed:2022qnc} topological boundary condition determines how the symmetry category is realized in the QFT. For example, if a QFT has a non-anomalous group-like discrete global symmetry $G$, the quiche boundary for the SymTFT admits boundary conditions which couple the QFT to background $G$-gauge fields as well as boundary conditions  which topologically gauge the $G$ symmetry (or its subgroups) in the QFT. 
The different boundary conditions for the SymTFT can be described as states in the SymTFT Hilbert space when we quantize along $M_d$ with $t$ as the time direction. We will denote the  choice of quiche boundary state as $|\mathfrak{B}\rangle$. 

The QFT boundary is more subtle, as it must encode the information of the partition function of a generically non-topological theory. Given an orthonormal basis of the SymTFT Hilbert space $|A\rangle$, the QFT boundary state can be regarded as 
\eq{
\langle QFT|=\sum_A Z_{QFT}[A]~\langle A|~,
}
where here $Z_{QFT}[A]$ is the partition function of the state realized in the phase with $|\fB\rangle=|A\rangle$. In general, the boundary conditions $|\fB\rangle$ are a linear combination of such states which allows for a broader notion of the realization of the symmetry category such as by incorporating discrete gaugings. The  framework of the SymTFT then naturally incorporates all possible realizations of the symmetry category on a QFT by realizing the QFT as a relative theory where the partition function is really a state in the Hilbert space of all possible quiche boundary conditions $\fB$. Let us illustrate this with an example. 
\vspace{0.25cm}

Consider the case where a $d$-dimensional QFT has a $\IZ_N^{(0)}$ 0-form global symmetry. Here, the SymTFT is $(d+1)$-dimensional $\IZ_N$ BF theory:
\eq{
S=\frac{iN}{2\pi}\int da_1\wedge b_{d-1}~,
}
where $a_1,b_{d-1}$ are degree-1, and degree-$(d-1)$ $U(1)$-valued gauge fields respectively. This theory has a pair of topological operators
\eq{
W_n(\gamma)=e^{i n \oint_\gamma a_1}\quad, \quad U_m(\Sigma)=e^{ i m \oint_\Sigma b_{d-1}}~,
}
which generate $\IZ_N^{(d-1)}\times \IZ_N^{(1)}$ symmetries of the SymTFT respectively and correspondingly have the non-trivial  correlation functions
\eq{
\langle W_n(\gamma)\, U_m(\Sigma)\rangle=e^{\frac{2\pi i n m}{N}{\rm Link}(\gamma,\Sigma)}~. 
}

For this theory, we can choose an orthonormal basis of  states  where the $a_1$ has Dirichlet boundary conditions $|A\rangle$ where the boundary value of $a_1$ is a flat $\IZ_N$ gauge field $a_1\big{|}_{bnd}=\frac{2\pi}{N}A$. In this basis of boundary states/conditions, the $a_1$ Wilson lines take definite values while the $b_{d-1}$-surface operators shift the boundary gauge field $A$ 
\eq{
e^{ i m \oint_\Sigma b_{d-1}}|A\rangle=\left|A+m\,\Omega(\Sigma)\right\rangle~,
}
where $\Omega(\Sigma)$ is the global angular form of $\Sigma$ which satisfies $d\Omega(\Sigma)=\delta(\Sigma)$.\footnote{Here we use the notation that $\delta(\Sigma)$ is the Thom class of the normal bundle of $\Sigma$. See for example \cite{Harvey:2005it}.}
In this basis, it is natural to expand the QFT state as
\eq{
\langle QFT|=\frac{1}{\sqrt{|H^1(M_d;\IZ_N)|}}\sum_{A\in H^1(M_d;\IZ_N)} Z_{QFT}[A]\,\langle A|~, 
}
where $Z_{QFT}[A]$ is the partition function of the QFT where we have coupled to a fixed background gauge field $A$ for the $\IZ_N^{(0)}$ global symmetry and $\langle QFT|A\rangle=Z_{QFT}[A]$. 

The SymTFT also admits 
Neumann boundary conditions for $a_1$ (which corresponds to Dirichlet boundary conditions for $b_{d-1})$  which can be written as 
\eq{
|N_{B}\rangle=\frac{1}{\sqrt{|H^1|}}\sum_{A\in H^1(M;\IZ_N)}e^{\frac{2\pi i }{N}\int_{M_d} A\cup  B}|A\rangle~,
}
where $B$ is a fixed $\IZ_N^{(d-2)}$ background gauge field. This boundary condition clearly corresponds to the case where the $\IZ_N^{(0)}$ global symmetry of the Dirichlet basis is gauged.

\section{Generalized Families of Theories}

\label{sec:genfam}

Recently, many of the concepts implemented in spurion analysis have also been revisited in the context of generalized/categorical global symmetries. 
The general idea of spurion analysis is the following.  Consider a QFT with a 0-form global symmetry $G^{(0)}$ which can be broken by adding some local operator/interaction to the theory:
\eq{
\Delta S=\sum_i\lambda^i \int d^dx~ \CO_i(x)+c.c.~. 
}
The coupling constants $\lambda^i\in \IC^N$ are dimensionful parameters according to the scaling dimension of $\CO_i(x)$. 
When the $\CO_i(x)$ transform non-trivially under $G^{(0)}$, we can additionally assign the $\lambda^i$ a $G^{(0)}$ transformation property so that the combination $\sum_i\lambda^i\, \CO_i(x)$ is $G^{(0)}$-invariant and the symmetry is artificially restored. Coupling constants may have natural transformation properties in physical systems when the coupling constant is set by the expectation value of some scalar field -- in which case $G^{(0)}$ is  spontaneously broken instead of explicitly broken. However natural the transformation of $\lambda^i$ is, 
endowing $\lambda^i$ with a $G^{(0)}$ transformation rule allows one to use many of the standard techniques of global symmetries to analyze the theory and how it depends on the value of $\lambda^i$.

The consequence of explicitly breaking a $G^{(0)}$ global symmetry in this way is that the deformed theory belongs to a family of theories which are $G^{(0)}$-equivariant -- i.e. a family of theories which is indexed by  some parameter $\lambda^i$ that takes values in a parameter space with a natural $G^{(0)}$ action. Because the $G^{(0)}$ action is a broken symmetry of the theory, moving along the $G^{(0)}$ orbits in parameter space are exactly marginal deformations when $G^{(0)}$ is continuous which we parametrize by dimensionless angles $\theta$; in terms of spurions, they are would-be Nambu-Goldstone bosons.   

The equivariant family structure of the theory admits a projective realization on the Hilbert space, and in particular on the  partition function. Consider a family of $d$-dimensional quantum field theories $\CT_\theta$ on $M_d$ indexed by the continuous exactly marginal couplings $\theta\in \CX_\theta$. Generically the couplings are valued in some space $\CX_\theta$ (which may be non-compact) and the partition function is  naturally a complex function on $\CX_\theta$. 
If the theories $\CT_\theta$ and $\CT_{\theta+\Lambda}$ can be identified,\footnote{Here we will not refer to this identification between different $\CT_\theta$ as a symmetry (although it is sometimes referred to as a $(-1)$-form global symmetry \cite{Aloni:2024jpb,Vandermeulen:2022edk,Vandermeulen:2023smx,Sharpe:2022ene,Yu:2024jtk}).} we can reduce the space of couplings by this identification to the reduced space of coupling constants $\CC_\theta=\CX_\theta/\Lambda$. The partition function $Z$ then generically descends to a complex line bundle over the reduced space $\CC_\theta$:
\eq{
\xymatrix{\IC\ar[r]&Z\ar[d]\\
&\CC_\theta=\CX_\theta/\Lambda}
}
For a family of $G^{(0)}$-equivariant theories, the relations generated by $\Lambda$ generally include the $G^{(0)}$ action on $\theta\in \CX_\theta$, which can be continuous or discrete. Notice that singular points of $\CC_\theta$ that arise from the $G^{(0)}$ action on $\CX_\theta$ correspond to symmetry enhancement points in parameter space. 

When there are additional (higher-form or higher-group) global symmetries $\CG$, we couple the partition function to background gauge fields. The partition function then becomes a line bundle over the space of coupling constants times background gauge fields:\footnote{Here we use the notation where $\CG=\prod_i G^{(p_i)}\times \prod_j \IG_{n_j}$ and $B_\nabla \CG=\prod_i B^{p_i+1}_\nabla G^{(p_i)}\times \prod_j B_\nabla\IG_{n_j}$ where $B^k_\nabla G$ is the space of connections on a $G$-valued $k$-bundle which are topologically classified by maps into $B^kG$.
}
\eq{
\xymatrix{\IC\ar[r]&Z\ar[d]\\&\CC_\theta\times B_\nabla \CG}
} 

A simple example that illustrates the structure of a continuous family of theories is $4d$ $SU(N)$ Yang-Mills theory with theta angle:
\eq{
S_\theta=...+\frac{i\theta}{8\pi^2}\int_{M_d} \Tr[F\wedge F]~. 
}
The theta angle $\theta \in \CC_\theta=S^1$ is $2\pi$-periodic, $e^{-S_\theta}=e^{-S_{\theta+2\pi}}$, since for $SU(N)$ the instanton number is an integer:
\eq{
\oint_{M_d} \frac{\Tr[F\wedge F]}{8\pi^2}\in \IZ~. 
}
Shifting $\theta\mapsto \theta+2\pi$ has a non-trivial effect due to the Witten effect \cite{Witten:1979ey} so the isomorphism 
\eq{
\CT_\theta\cong \CT_{\theta+2\pi}~
}
is non-trivial, e.g. it requires a redefinition of the electromagnetic charge lattice.  See for example~\cite{Chen:2025buv,Chen:2024tsx} for a recent discussion on aspects of $\theta$ terms.

\subsection{Higher Group Structures in the Space of Coupling Constants}
 \label{sec:higherfam}
 
We will consider the structure that arises when more general types of global symmetries are explicitly broken. We first consider explicitly breaking a higher group global symmetry $\IG_n$ by adding a local interaction which breaks the  
$n$-group down to a sub-$n$-group: $\IG_n\to \IG_n^{(1)}$  by breaking the 0-form component of $\IG_n$.   
The spurion paradigm then generalizes by assigning a transformation property to the symmetry breaking coupling constant that additionally requires a transformation of the background gauge fields for the unbroken $\IG_n^{(1)}$. The broken $\IG_n$ action on the coupling constant will lead to an identification which requires a shift of both the coupling constant and the $\IG_n^{(1)}$ background gauge fields. 

To consider explicit breaking of the 0-form part of a higher group global symmetry,
recall the background gauge transformation of the gauge fields of the higher group in \eqref{generalHGtrans}
\eq{
\delta A_{p+1}=d\lambda_p+\sum_{q\leq p}\lambda_q\wedge \omega_{p-q+1}(A,\lambda)~.
}
Notice that breaking the 0-form component can trivialize the higher group structure. The reason is that in general, smoothly varying spatially dependent coupling constants only correspond to flat gauge fields for the broken symmetry, 
similar to the case when there are Goldstone fields for a spontaneously broken symmetry. Thus, when the $\omega_k(A,\lambda)$ vanish for flat 0-form background gauge fields $A_1$  
the corresponding symmetry structure is trivialized. 

For example, consider explicitly breaking a continuous 2-group in a  $4d$ QFT which arises from perturbative cubic anomalies of $U(1)$ 0-form global symmetries similar to the example described in Section \ref{sec:highergroup}. The 2-group structure will generically be trivializable for any component involving the broken 0-form symmetry due to the fact that the relevant transformations of $B_2$ in \eqref{2grouptransexample} depend on the field strength of the explicitly broken 0-form global symmetry. 

To avoid trivializing the higher symmetry, we consider cases where the higher group global symmetry contains a discrete sub-2-group that  
mixes a 0-form internal abelian symmetry and 0-form Lorentz symmetry into a higher group. 
For example,  consider a family of theories indexed by $\theta$ such that shifting the parameter $\theta\mapsto \theta+2\pi$ causes the action to shift by 
\eq{\label{dummyeq}
\Delta S=\pi i \int W_{p+1}\wedge \ast J_{p+1}~, 
}
where $W_{p+1}$ is a choice of integral lift of a $\IZ_2$-valued degree $(p+1)$ characteristic class that only depends on the Stiefel-Whitney classes of spacetime and 
$\ast J_{p+1}$ is a conserved current or otherwise quantized dynamical field-dependent operator. In this example, $\ast J_{p+1}$ acts as a current operator which couples to a background gauge field $B_{p+1}$ as
\eq{
S=...+i \int B_{p+1}\wedge \ast J_{p+1}~.
}
Then, the shift in the action in \eqref{dummyeq} can be absorbed by shifting $B_{p+1}\mapsto B_{p+1}+\pi W_{p+1}$. This leads to 
an identification between the theory with background gauge fields
\eq{\label{higherfamIdentex}
\CT_{\theta}[B_{p+1}]\cong  \CT_{\theta+2\pi}[B_{p+1}+\pi W_{p+1}]
~\Longrightarrow~Z_\theta[B_{p+1}]\sim Z_{\theta+2\pi}[B_{p+1}+\pi W_{p+1}]~. 
}

Using background gauge fields to cancel a variation of the action is reminiscent of the Green-Schwarz mechanism and the idea of higher group global symmetries  \cite{Cordova:2018cvg,Brennan:2020ehu}. Because of this, we refer to these families of theories as having \emph{higher group structure in the space of coupling constants} or that the family is a  \emph{higher family} of QFTs.

This higher family structure also naturally extends to a mathematical structure on the partition function. Recall from our previous discussion that the partition function  of a  family of theories with $\CG$ symmetry forms a line bundle over the space of coupling constants and space of background gauge fields $\CC_\theta\times B_\nabla \CG$:  
\eq{
\xymatrix{\IC\ar[r]&Z\ar[d]\\
&\CB=\CC_\theta\times B_\nabla \CG}
}
when there is an identification $\CT_\theta\cong \CT_{\theta+2\pi}$. 
When there are additional identifications from the higher family structure of the form in \eqref{higherfamIdentex}, 
we can further reduce the parameter space:
\eq{\label{highergrouplinebundle}
\xymatrix{\IC\ar[r]&Z\ar[d]\\
&\CB/\Lambda=\frac{\CC_\theta\times B_\nabla \CG}{\Lambda}
}
}
where $\Lambda$ are the higher family relations that come from broken $\IG_n$-generators.  
Again, the partition function can be a non-trivial line bundle over the reduced space $\CB/\Lambda$. In analogy with ordinary continuous families of theories, the partition function descending to a non-trivial line bundle over $\CB/\Lambda$ should be interpreted as the theory having 
`t Hooft and generalized family anomalies as appropriate. We will discuss this further in the next section.

\subsubsection{Constraints on RG Flows}
One question we can ask is if there are any physical implications inherent to the higher family structure. 
 To answer this, recall that when higher group global symmetries are accidental symmetries of the IR theory, there must be a hierarchy of energy scales along which the symmetry emerges. Since higher groups have a natural grading
\eq{
\IG_n\supseteq \IG_n^{(1)}\supseteq \IG_n^{(2)}\supseteq...\supseteq \IG_n^{(n-1)}=G^{(n-1)}~,
}
where $\IG_n^{(p)}$ is the $n$-subgroup that contains only $q\geq p$-form components, $\IG_n$ global symmetry can only emerge in steps that respect the nested structure:
\eq{
E_{\IG_n}\lesssim E_{\IG_n^{(1)}}\lesssim E_{\IG_n^{(2)}}\lesssim...\lesssim E_{G^{(n-1)}}~,
}
where $E_{\IG_n^{(p)}}$ is the energy scale where the $\IG_n^{(p)}$ $n$-group global symmetry emerges. 

Similarly, given a higher family of theories, we can ask if the higher family structure places constraints on RG flows or on the possible UV completions. Consider a family of theories which is parametrized by $\theta$ which is periodic $\theta\sim \theta+2\pi$ and let us assume that in the IR, the family of theories has a $\IG_n^{(1)}$ global symmetry that  participates in an emergent higher family structure (possibly modulo an anomalous phase):
\eq{
\CT_\theta[B]\cong \CT_{\theta+\frac{2\pi}{N}}[B+\Lambda_{fam}]\quad \Longrightarrow \quad Z_\theta[B]=Z_{\theta+\frac{2\pi}{N}}[B+\Lambda_{fam}]
}
where $B$ are the background gauge fields for $\IG_n^{(1)}$ and $N\geq2$ is a positive integer.  For this family of theories, we can distinguish between the UV behavior of $\IG_n^{(1)}$ -- i.e. whether or not the global symmetry is preserved or broken in the UV theory -- by checking if the partition function is invariant under $\IG_n^{(1)}$ background gauge transformations at all energy scales $E$:\footnote{If $\IG_n^{(1)}$ is anomalous in the IR, the relation in \eqref{higherfamE} would be modified by an anomalous phase.}
\eq{\label{higherfamE}
Z_\theta^{(E)}[B+d\Lambda_\IG]\stackrel{?}{=}Z_\theta^{(E)}[B]~.
}
Similarly, if we restrict to UV families of theories parametrized by $\theta\sim \theta+2\pi$,\footnote{Here we only consider UV completions where the family parametrized by $\CC_\theta$ holds along the entire RG flow. One could more generally consider the case where the family relations that give rise to $\pi_1(\CC_\theta)\neq 0$ are also only properties of the IR theory, but we will not discuss this scenario here.}  we can also determine if the higher family structure is violated at some energy scale $E$ by checking
\eq{
Z_{\theta+\frac{2\pi}{N}}^{(E)}[B+\Lambda_{fam}]\stackrel{?}{=}Z_\theta^{(E)}[B]
}
Since the family structure can activate a $\IG_n^{(1)}$ background gauge transformation, it is inconsistent for a theory to have the higher family structure without the $\IG_n^{(1)}$ global symmetry. So the only possible UV behaviors of a family of theories with IR higher family structure are:\footnote{If we allow for higher dimensional UV completions, then the higher family structure can also be matched by a genuine higher group global symmetry. We will discuss the matching of the higher family structure by a higher group global symmetry in Section \ref{sec:compactification}.  We do not attempt to classify the possible higher dimensional UV completions.}

\begin{enumerate}
\item The higher family structure is preserved along the RG flow $(\IG_n^{(1)}$ is preserved), 
\item There is $\IG^{(1)}_n$ symmetry in the UV and the higher family structure is emergent along the RG flow, or
\item There is no $\IG^{(1)}_n$ symmetry in the UV and the higher family structure is emergent along the RG flow.
\end{enumerate}

Let us comment on the cases where the higher family structure is IR-emergent along the RG flow.  Families where $\IG_n^{(1)}$ is a symmetry of the UV theory, but the family structure is enhanced to a higher family structure below some energy scale $E_{\rm fam}$,  
can arise if the operator multiplying $\theta$ in the IR is mixed with another dimension-$d$ operator in the UV 
whose shift cannot be canceled by a compensating $\IG_n^{(1)}$ transformation. This occurs for example in $4d$ non-abelian completions of abelian gauge theory, with $E_{\rm fam}$ set by the mass of the W-bosons at weak coupling, or mass of the monopoles at strong coupling.  These scales set the scale of the non-abelian structure/size of the instanton core which breaks the higher family structure. 
 
The other way to have emergent higher family structure is if the $\IG_n^{(1)}$ global symmetry that participates in the higher family structure is itself emergent at some scale $E_{\IG^{(1)}}$. In this case, at energies $E\gg E_{\IG^{(1)}}$, the partition function cannot be invariant under the higher family transformation because the family relations close onto the $\IG^{(1)}_n$ background gauge transformations plus the inherent $\CC_\theta$ periodicity.  
This implies that emergent higher family structure comes with the (parametric) bound on energy scales:
\eq{
E_{\IG^{(1)}}\gtrsim E_{\rm fam}
} 
 similar to the case of higher group global symmetries.

\subsubsection*{Example: $4d$ Abelian Gauge Theory and its possible UV completions}

Consider $U(1)$ Maxwell theory; this discussion can be straightforwardly generalized to a $U(1)^r$ gauge theory, as can arise in the Coulomb phase of a UV non-Abelian gauge theory, with 
\eq{
S_\theta=\frac{i}{8\pi^2}\int \theta\,F\wedge F~. 
}
The $S_\theta$ term has $\theta \sim \theta +2\pi$ periodicity if the spacetime is spinnable ($w_2(TM)=0$) and generally only $\theta \sim \theta +4\pi$ periodicity if not ($w_2(TM)\neq 0$) with 
\eq{
S_{\theta+2\pi n }-S_{\theta}=i \pi  n\int w_2(TM)\cup \frac{F}{2\pi}~. 
}
This anomaly can be rephrased as a 2-group, with the $\Delta \theta =2\pi$ shift canceled by a compensating shift in the $U(1)^{(1)}_m$ background:
\eq{
\Delta \theta =2\pi n \quad, \quad \Delta B_2=n \pi w_2(TM)~. 
}

If this Maxwell theory is UV completed into the Cartan of a $G_{UV}$ non-Abelian gauge theory, the IR $\theta$ here maps to the UV theta angle $\theta_{UV}$, which may have a similar higher family relation depending on the global form of $G_{UV}$ provided there is a $\IZ_2^{(1)}$ global symmetry in the UV theory so that
\eq{
\CT_{\theta_{UV}}[B]\cong \CT_{\theta_{UV}}[B+\pi w_2(TM)]~,
}
where $B$ is the $\IZ_2^{(1)}$ background gauge field  
 that cancels the shift of $\theta_{UV}$.  

Since different gauge groups can have different fractional instantons, there are different possible scenarios for the different UV gauge groups \cite{Aharony:2013hda,Witten:2000nv}. If we match the minimal instanton of the UV non-abelian theory to that of the IR Coulomb theory, we see that if the UV gauge group is  $SU(N),PSU(2N+1),Spin(N),Sp(N),E_6,E_7$, then the $\IZ_2^{(1)}\subset U(1)_m^{(1)}$ is broken.  The  higher family identification must then also be broken, with  $E_{\text{1-form}}\sim E_{fam}$.  

On the other hand, if the UV gauge group is $PSU(2N),SO(2N),PSp(2N+1),E_7/\IZ_2$, the 1-form symmetry need not be spontaneously broken.  The fractional instantons of the UV theory then lead to a higher family of theories in the UV, with the family identification preserved along the entire RG flow. 

For the UV completions with gauge group $SO(2N+1)$ and $PSp(2N)$ for $N>1$ there is an unbroken $\IZ_2^{(1)}$ 1-form magnetic symmetry, but there are no fractional instantons so that there is no higher family structure in the UV. This is because the $\IZ_2^{(1)}$ center symmetry of $Spin(2N+1)$ and $Sp(2N)$ embeds into the electric $(U(1)^{(1)})^N$ 1-form symmetry of the IR theory so that the $\IZ_2$ fluxes of the $SO(2N+1)$ and $PSp(2N)$ gauge theory do not activate a fractional instanton in the IR. In other words, the $\IZ_2^{(1)}\subset U(1)^{(1)}_m$ that participates in the higher family structure in the IR is not the UV $\IZ_2^{(1)}$ 1-form magnetic symmetry.

\subsection{Categorical Structures in the Space of Coupling Constants}
\label{sec:NonInvertible}

We can also consider what structure arises from explicitly breaking non-invertible global symmetries. 
We will restrict our attention to symmetries that arise from discrete gauging such as 
non-invertible chiral symmetries in $4d$ theories -- e.g. subgroups of the axial $U(1)_A^{(0)}$ global symmetries that are transmuted to a non-invertible symmetry by the ABJ anomaly as in~\cite{Choi:2022jqy,Cordova:2022ieu}.  The explicitly broken non-invertible global symmetry can still be useful if we assign the symmetry violating coupling constant $\theta$ with an appropriate transformation property. This leads to a family of theories which are identified by the (broken) non-invertible symmetry transformation where the shift of the coupling constant $\theta\mapsto \theta+\theta_0$ is accompanied by a discrete gauging of the relevant symmetry. Spurion analysis for broken non-invertible symmetries has been used in the context of particle physics in e.g. \cite{Cordova:2022qtz,Cordova:2024ypu,Choi:2025vxr}. 

To illustrate the idea, consider the example of $4d$ QED with a single Dirac fermion, with relevant deformation by a mass term:
\eq{
\CL=\frac{1}{2g^2}F\wedge \ast F+i \bar\psi \slashed{D}\psi+m \bar\psi \psi+c.c.
}
where $m\in \IC$.   As reviewed in Section \ref{sec:review}, when $m=0$ this theory has an (anomalous) $U(1)$ chiral symmetry which is transmuted to a $(\IQ/\IZ)^{(0)}$ non-invertible global symmetry.  The non-invertible $q=p/N\in \IQ/\IZ$ symmetry operator (where ${\rm gcd}(p,N)=1$) can be realized in the half-gauging formalism as a domain wall between the theory $\CT$ and $\CT/\IZ_N^{(1)}$: the theory where we have gauged a $\IZ_N^{(1)}\subset U(1)^{(1)}_m$ global symmetry with a torsion term \cite{Choi:2021kmx,Choi:2022zal,Roumpedakis:2022aik}.  We refer to this gauging operation as $\CS_q$ for $q=\frac{p}{N}\in \IQ/\IZ$ with $p,N$ co-prime. Because $(\IQ/\IZ)^{(0)}$ is a symmetry, the half-gauging domain wall is topological and we have the relation 
\eq{
\CT\cong \CS_q\CT~.
}
Consequently, we can compare the partition functions of $\CT$ 
\eq{
Z_\CT&=
\int [dA_1]~Z_{\CT}[A_1]~,
}
with the partition function of $\CS_q\CT$:
\eq{
 Z_{\CS_q\CT}&=\int [dA_1]Z_{\CS_q\CT}[A_1]= \int [dA_1dc_1db_2]~Z_{\CT}[A_1]\times e^{ \frac{i N}{2\pi}\int b_2\wedge dc_1+i \int \frac{F_2}{2\pi}\wedge b_2+\frac{i N k}{2\pi}\int\frac{\CP(b_2)}{2}}~,
}
where $k$ even for $N$ odd, $kp=1$ mod$_{\gamma(N)N}$,  $\CP$ is the Pontryagin square,\footnote{For $N$ odd, the Pontryagin square is simply $\CP(b_2)=b_2\cup b_2$. 
} $F_2=dA_1$, and 
\eq{
\gamma(N)=\begin{cases}
1&N\text{ odd}\\
2&N\text{ even}
\end{cases}
}
Since $\CT\cong \CS_q\CT$, these partition function must be equivalent up to a possible phase.

Following \cite{Apte:2022xtu,Choi:2022zal}, we now demonstrate that $Z_\CT[A_1]$ and $Z_{\CS_q\CT}[A_1]$ are indeed equivalent up to such a phase.  First integrating out $c_1$ restricts the path integral to a discrete sum over $b_2\in H^2(M;\IZ_N)$
\eq{
 Z_{\CS_q\CT}[A_1]&=\frac{Z_{\CT}[A_1]}{\sqrt{|H^2(M;\IZ_N)|}} \sum_{b_2\in H^2(M;\IZ_N)} e^{\frac{2\pi i k}{2N}\int \CP(b_2) +\frac{2\pi i}{N}\int \frac{F_2}{2\pi}\cup b_2}~.
}
As discussed in \cite{Apte:2022xtu,Choi:2022zal}, $Z_{\CS_q\CT}[A_1]$ can be computed by evaluating the Gauss sum:
\eq{
Z_{\CS_q\CT}[A_1]=Z_{\CT}[A_1]\times G(M,p,N)\times e^{-\frac{2\pi i p }{2 N}\int \frac{F_2\wedge F_2}{4\pi^2}}~, 
}
which is indeed equivalent to $Z_{\CT}[A_1]$ up to a phase. The phase is given in terms of $p,N$ and  
the topology of spacetime through 
\eq{
G(M,p,N)=\begin{cases}\left(\epsilon_N \left(\frac{k/2}{N}\right)\right)^{\sigma(M)}&\text{odd N}\\
\left(e^{\frac{\pi i }{4}}\epsilon_k^{-1}\left(\frac{2N}{k}\right)\right)^{\sigma(M)}&\text{even N}
\end{cases}\quad, \quad pk=1~{\rm mod}_{\gamma(N) N}~,
}
where $\sigma(M)$ is the signature of $M$ and 
\eq{
\epsilon_N=\begin{cases}
1&N=1~{\rm mod}_4\\
i &N=3~{\rm mod}_4
\end{cases}\quad, \quad \left(\frac{a}{s}\right)=\begin{cases}
1&s=1\\
0&s\neq 1~,~a=0~{\rm mod}_s \\
1&s\neq 1~,~a\neq 0~{\rm mod}_s\text{ and }a=n^2~{\rm mod}_s\\
-1&\text{else}
\end{cases}
}

Now, because the $U(1)$ chiral symmetry has a $\Tr U(1)$ gravitational anomaly, there is a field redefinition (i.e. chiral rotation) of the fermions in the path integral to absorb the $F_2$-dependent phase of $Z_{\CS_q\CT}[A_1]$ at the cost of an additional gravitational phase:\footnote{The $U(1)$ chiral symmetry of a Dirac fermion has a mixed (perturbative) gravitational anomaly with the same anomaly coefficient as the ABJ anomaly that gives rise to the phase factor in \eqref{gravphaseCSq}. 
This follows from the Atiyah-Singer Index theorem for a fermion coupled to a $U(1)$ gauge field:
\eq{
{\rm Ind}[\slashed{D}]=\int \left(\frac{c_1^2}{2}-\frac{p_1}{24}\right)=\half\int \left(c_1^2-\frac{p_1}{12}\right)~.
}
} 
\eq{\label{gravphaseCSq}
Z_{\CS_q\CT}[A_1]=Z_\CT[A_1]\times G(M,p,N)\times e^{-\frac{2\pi i p}{2 N}\int \frac{p_1(M)}{12}}~.
}
Although $(\IQ/\IZ)^{(0)}$ is a non-invertible symmetry, the relation of theories is in a sense ``invertible'' in that the partition functions are equal up to a phase:
\eq{
\CT\cong \CS_{-q}\CS_q\CT\quad \Longleftrightarrow \quad \big{|}Z_\CT\big{|}=\big{|}Z_{\CS_{-q}\CS_q\CT}\big{|}~.
}

Let us demonstrate this for $q=1/N$ where $N$ is odd. The transformed partition function is given by
\eq{
Z_{\CS_{-q}\CS_q\CT}[A]=\int [dc\,d\tildec \,db\,d\tildeb]Z_{\CT}[A]\times e^{\frac{2\pi i }{N}\int (b\cup \delta c+\tildeb\cup \delta \tildec)+\frac{2\pi i }{N}\int (b\cup b-\tildeb\cup \tildeb)+\frac{2\pi i}{N}\int \frac{F}{2\pi}\cup (b+\tildeb)}
}
If we integrate out $c,\tildec$, then we get the finite sum:
\eq{
Z_{\CS_{-q}\CS_q\CT}[A]&=\frac{Z_{\CT}[A]}{{|H^2(M;\IZ_N)|}} \sum_{b,\tildeb\in H^2(M;\IZ_N)} e^{\frac{2\pi i }{N}\int (b\cup b-\tildeb\cup \tildeb+\frac{F}{2\pi}\cup(b+\tildeb))}
}
We can then reparametrize $b,\tildeb$ as $b_\pm =b\pm \tildeb$, \footnote{This reparametrization requires $N$ odd. To see this, consider a generator $[\Sigma]\in H_2(M;\IZ)$ and let $\oint_\Sigma b_-=n$ be fixed. The periods $\oint_\Sigma b_+$ then take values $ [n,n+2,...,n+2N-2]\sim [0,1,...,N-1]$ mod$_N$ if $N$ is odd so that  the sum over $b_\pm$ can be straightforwardly  implemented as above. }  
so that $Z_{\CS_{-q}\CS_q\CT}[A]$ is given by 
\eq{
Z_{\CS_{-q}\CS_q\CT}[A]&=\frac{Z_{\CT}[A]}{{|H^2(M;\IZ_N)|}} \sum_{b_\pm \in H^2(M;\IZ_N)} e^{\frac{2\pi i }{N}\int (b_+\cup b_--\frac{F}{2\pi}\cup b_+)}~.
}
Computing the finite sum over $b_+$ then implements a Kronecker delta function 
\eq{
Z_{\CS_{-q}\CS_q\CT}[A]&=Z_{\CT}[A]\sum_{b_-\in H^2(M;\IZ_N)} \delta\left(b_-+\frac{F}{2\pi}~{\rm mod}_N\right)=Z_{\CT}[A]~. 
}
 
When $m\neq 0$, the $\IQ/\IZ$ non-invertible symmetry is broken. We can then assign the mass term the transformation $m\mapsto \widetilde{m}=e^{-4\pi i q} m$ for $q\in \IQ/\IZ$ to preserve the global symmetry.   
The (broken) symmetry transformation then ensures that the two theories are identical:
\eq{
\CT_m\cong \CS_q \CT_{\widetilde{m}}~. 
} 
This allows us to compare the two partition functions $Z_m,\CS_q Z_{\widetilde{m}}$ and we see that as in the case with group-like broken symmetries, the partition functions must be equivalent up to a possible phase:
\eq{
Z_{m}=\CS_q Z_{ \widetilde{m}}\times e^{ i \int \omega}~. 
}
This identity follows from the computation of $|Z_\CT|=|Z_{\CS_q \CT}|$ as in the massless case above.

Such family structure phenomena arise for any theory that is symmetric under a discrete gauging that can be broken by deforming the action by a local operator with coupling constant parametrized by a dimensionless parameter $\theta$. 
Again we denote the gauging action by $\CS$ and we find that the relation  $\CT_\theta\cong \CS\CT_{\theta'}$ 
implies the relation 
\eq{
|Z_{\theta}|=|\CS\,Z_{\theta'}|~. 
}
 This occurs more generally when there is a 0-form global symmetry that has an abelian ABJ anomaly, but can also occur for example in $2d$ theories with non-invertible Kramers-Wannier-like symmetries \cite{Chang:2018iay,Kaidi:2021xfk,Frohlich:2004ef}.

We will interpret the broken non-invertible symmetry structure as a generalized type of family identification.  
In order to interpret the $\CS$ gauging operation as relating different theories in the same family, we need to expand our notion of what constitutes a ``family of QFTs.''  One natural way to expand our notion of a family of theories is to consider theories that can be related by a non-trivial interface to this theory. Of course, the family of all such theories is too broad since any theory that admits symmetry preserving boundary conditions to the trivial theory can be related to any other theory with similar properties. To realize explicitly broken non-invertible symmetry as a family structure, we will extend the notion of a family of theories to include the space of theories that are related by different topological manipulations.   
 This extends the notion of a family of theories by the space of topological (quiche) boundary conditions allowed in the theory's SymTFT. 

In the framework of the SymTFT, it is natural to consider a family of theories that is defined by the QFT state for a theory $\CT_\theta$:
\eq{
\langle \CT_\theta|=\sum_{A}Z_{\theta}[A]\langle A|~,
}
which allows for parameters in the theory (here denoted by $\theta$) as well as all possible inner products with the quiche boundary state $|B\rangle$. We can project to a particular element in the generalized family of theories by taking the inner product of this ``QFT family'' state with some topological boundary condition:
\eq{
Z_{\theta}[B]=\langle \CT_\theta|B\rangle~. 
} 
The partition function is then a line bundle above the space of parameters together with the space of  possible quiche boundary conditions $\mathfrak{B}$:
\eq{ 
\xymatrix{\IC\ar[r]&Z\ar[d]\\&\CB=\CC_\theta\times \mathfrak{B}}
}
The space of topological boundary conditions $\fB$ is given by the Hilbert space for the SymTFT and hence the family of partition functions is a line bundle over the space $\CC_\theta\times \fB$ which is finite dimensional for finite symmetries. 
For the case of group-like global symmetries $\CG$, the set of boundary conditions $\fB$ contains the space of $\CG$ gauge connections  $B_\nabla \CG\subset \mathfrak{B}$, thereby including the notion required for higher family structures from the previous section. The set $\mathfrak{B}$ also includes other components, such as all possible topological gaugings.

In this language, the action of $\CS$ implements a (discrete) shift along $\mathfrak{B}$. Explicitly, consider a family $\CT_\theta$  of $4d$ gauge theories with a $\theta$-term and a $\IZ_N^{(1)}$ global symmetry that can be gauged to produce the non-invertible family relation (e.g. via explicitly breaking the $\IZ_N\subset \IQ/\IZ$ non-invertible symmetry in $4d$ QED). 
We denote the Dirichlet boundary condition for the $\IZ_N^{(1)}$ symmetry as $|B\rangle$.  The $\CS_q$ operation gauges the 1-form symmetry, along with a $q$-dependent torsion term,  
\eq{
\CS_q:|B\rangle \longmapsto \frac{1}{\sqrt{|H^2(M;\IZ_N)|}}\sum_{b\in H^2(M;\frac{2\pi}{N}\IZ_N)} e^{\frac{i kN}{4\pi}\int \CP(b-B)}|b\rangle~,
}
where $q=\frac{p}{N}$ for $p,N$ co-prime and $ pk=1$ mod$_{\gamma(N) N}$ with $k$ even for $N$ odd. The inner product of $\CS_q|B\rangle$  with $\langle \CT_\theta|$ is 
\eq{
\CS_q Z_{\theta}[B]&=\langle \CT_\theta|\CS_q|B\rangle=\int [dc_1\,db_2\,dA] \,Z_{\theta}[A]\, e^{\frac{iN}{2\pi}\int \left(dc_1\wedge b_2+\frac{k}{2}\CP(b_2-B)+\frac{1}{N}F\wedge b_2\right)}\\
&=\int [dA]\,Z_{\theta}[A]\,e^{\itwopi \int F_2\wedge B_2-\frac{2\pi i p}{ N}\int \frac{F\wedge F}{8\pi^2}}=\langle \CT_{\theta+2\pi q}|B\rangle=Z_{{\theta+2\pi q}}[B]~,
}
where here we are ignoring gravitational phases and we assumed   that
\eq{
Z_{\theta}[A]=Z_{{\theta+2\pi q}}[A]\times e^{2\pi i q \int \frac{F\wedge F}{8\pi^2}}~.
} 
It follows from the fact that $Z_{\theta}[B]=\CS_q Z_{\theta-2\pi q}[B]$ that $\CS_{-q}\circ \CS_q$ must act as the identity on $|B\rangle$.  Indeed we can easily check this (for simplicity we assume that $N$ is odd):
\eq{
\CS_{-q}\circ \CS_q|B\rangle
&=\frac{1}{|H^2|}\sum_{b,c} e^{\frac{i kN}{4\pi}\int (b-B)\cup (b-B)-\frac{ikN}{4\pi }\int (b-c)\cup(b-c)}|c\rangle\\
&=\frac{1}{|H^2|}\sum_{b,c} e^{\frac{i kN}{4\pi}\int 2b\cup(B-c)-c\cup c+B\cup B}|c\rangle\\
&=\sum_c \delta(c-B)\, e^{-\frac{i kN}{4\pi}\int c\cup c-B\cup B}|c\rangle=|B\rangle~,
}
so indeed $\CS_{-q}\CS_q Z_{\theta}[B]=Z_{\theta}[B]$.
As in the case of higher family structures, the partition function descends to a generally non-trivial bundle over the reduced parameter space 
\eq{\label{noninvertiblelinebundle}
\xymatrix{\IC\ar[r]&Z\ar[d]\\
&\CB/\Lambda=\frac{\CC_\theta\times \fB}{\Lambda}}
}
where $\Lambda$ are the discrete non-invertible shifts generated by $\CS_q$ together with $\theta\mapsto \theta-2\pi q$. 

This is a consistent, extended notion of families of QFTs, including categorical  symmetries. We refer to such a family of theories as having a \emph{categorical/non-invertible structure in the space of coupling constants} or as being a  \emph{categorical/non-invertible family of QFTs. }

\subsubsection{Constraints on RG Flows from (non-invertible) family structures}
Our examples of non-invertible family structures arise when an anomalous  phase in the path integral can be canceled by a discrete gauging of a higher form global symmetry $\CG$.  As in the case of higher family structures, we can consider an IR effective field theory with a non-invertible family structure and consider the possible UV completions.  For simplicity, we consider families of RG flows parametrized by a single parameter $\theta$ (the case of a family of theories parametrized by a more general space $\CC_\theta$ follows straight-forwardly), with periodicity $\theta\sim \theta+2\pi$.  We suppose that the IR theories have a global symmetry $\CG$ which participates in the non-invertible family structure.  We can again distinguish between the possible UV completions by tracking the global symmetry $\CG$ along the RG flow. Since the non-invertible family relation requires gauging the $\CG$ symmetry, UV completions with broken $\CG$ symmetry also cannot obey the non-invertible family structure. 
The possibilities for UV completions of such a non-invertible family of theories are:\footnote{We restrict ourselves to RG flows of theories that are in fixed spacetime dimension. In the case where we allow UV completions of higher dimension, there is also the possibility that the theory is completed by a $\CG$-preserving family of UV theories which participates in a genuine non-invertible global symmetry. This possibility is discussed in Section \ref{sec:compactification}. We do not claim to classify all possibilities with non-fixed dimension. }

\begin{enumerate}
\item The family of UV theories belongs to a $\CG$-preserving non-invertible family of theories that is preserved along the RG flow, 
\item The family of UV theories is $\CG$ preserving, but the non-invertible family identification is emergent in the IR, or
\item The family of UV theories does not have $\CG$ global symmetry or the non-invertible family relation; the entire structure is IR-emergent. 
\end{enumerate}
 
We first consider the case where the UV theory has a $\CG$ global symmetry and the categorical family structure is emergent in the IR, below some scale $E_{\rm fam}$.  The UV partition function can violate the enhanced IR structure if, for example,  
the operator that couples to the parameter $\theta$ mixes with another operator in the UV whose integral does not factorize modulo $N$ (as appropriate), spoiling the non-invertible family relation. This happens for example in gauge theories where fractional instantons are UV-completed into constrained instantons of a larger gauge group. In this case, the scale $E_{\rm fam}$ is set by the size of the instanton core above which the IR instanton density mixes into the instanton density of the larger gauge group which does not factorize modulo some integer $N$.

Now consider the case where $\CG$ is not a symmetry of the UV theory, but emerges at some scale $E_{\CG}$. As in the case of higher family structure, the fact that the non-invertible family structure requires the existence of the $\CG$ global symmetry (since it is gauged to produce the relation) means that there is a hierarchy of  scales 
\eq{
E_{\CG}\gtrsim E_{\rm fam}~. 
}

\subsubsection*{Example}

Consider the example of $4d$ $U(1)$ gauge theory with a $\theta$-angle
\eq{
S_\theta=-\frac{i\theta}{8\pi^2} \int F\wedge F~. 
}
We can consider a generalized family of theories with a non-invertible family relation 
\eq{
\CT_{\theta +2\pi q }\cong \CS_q \CT_{\theta }~, 
}
with $q \equiv p /N $ with ${\rm gcd}(p ,N )=1$.  The non-invertible $\CS_q$ operation gauges a $ \IZ_{N }^{(1)}\subset U(1)_m^{(1)}$:
\eq{
\CS_q\int [dA ]~Z_{\theta}[A ]~\longmapsto \int [dA ][dc ][db ]~Z_{\theta}[A ]\,e^{  \int \frac{iN }{2\pi}b \wedge dc +\frac{iN  k }{4\pi}b \wedge b +\frac{i}{2\pi}b \wedge F }
}
where $k  p =1$ mod$_{\gamma(N )N }$ and $k $ is even if $N $ is odd, the $A $ are  $U(1) $ gauge fields, and $c ,b $ are auxiliary 1- and 2-form $U(1)$ gauge fields to enact the gauging respectively. See \cite{Niro:2022ctq} for a related discussion on $SL(2,\IQ)$ non-invertible symmetries in Maxwell theory. 

This theory can be UV-completed into a non-abelian gauge theory. As in the case of the higher family, the completion to a UV gauge theory with Lie group  $G_{UV}$ (a finite subset of) the non-invertible family relations may be matched by a non-invertible family structure of the UV theory depending on the form of $G_{UV}$.   At the scale of non-abelian gauge symmetry restoration, we also find that the instantons generically break the $\IQ/\IZ$ family relations down to (at most) a discrete subgroup due to the UV instantons. Again, these examples satisfy the family hierarchy bound for the broken part of the non-invertible family relations and additionally when $G_{UV}$ includes fractional instantons, the UV completion belongs to a non-invertible family whose relations are preserved along the RG flow.  

\subsubsection{$1+1d$ Ising Model and Kramers-Wannier Duality}

The Ising model is a classic example of a categorical family of theories that arises from explicitly breaking a non-invertible symmetry involving gauging.  Kramers-Wannier duality can be phrased in terms of a $\IZ_2^{(0)}$ global symmetry that carries an ABJ anomaly ~\cite{Kapustin:2014gua, Shao:2023gho}.  In the continuum Fermion description, the Ising model is a free Majorana fermion coupled to a dynamical $\IZ_2$ gauge field $a$ that gauges fermion number. This is related via bosonization to the bosonic description of the Ising model. See e.g. \cite{Karch:2019lnn, Shao:2023gho} for reviews.  

The $\IZ_2$ gauge fields coupled to fermions can anomalously depend on the topology of the 2d (orientable, Euclidean) spacetime manifold $M$ via the  
Arf invariant on $H^1(M;\IZ_2)$, i.e. the mod-2 index of the Dirac operator $\CI$ coupled to the associated $\IZ_2$-gauge field:
\eq{
{\rm Arf}[a\cdot\rho]=\CI[a\cdot\rho]~.
}
Here $\rho$ is a base choice of spin structure and $a\cdot\rho$ is the spin structure where $\rho$ is twisted by the $\IZ_2$ gauge field $a$ and 
 the mod-2 index satisfies the condition 
 \eq{\label{mod2Identity1}
 \CI[(a+b)\cdot\rho]=\CI[a\cdot\rho]+\CI[b\cdot\rho]+\CI[\rho]+\int a\cup b~{\rm mod}_2~.
 }
For oriented manifold $M$, this implies 
 \eq{\label{mod2Identity2}
 \frac{1}{\sqrt{|H^1(M;\IZ_2)|}}\sum_a (-1)^{\CI[a\cdot\rho]+\CI[\rho]+\int a\cup b}=(-1)^{\CI[b\cdot \rho]}~. 
 }
 See \cite{Karch:2019lnn} for more details. The Kramers-Wannier transformation originates from the  $\IZ_2^c$ chiral transformation in the Majorana fermion description:
 \eq{
 \chi\longmapsto i \gamma_3\chi~. 
 }
The $\IZ_2^c$ has an ABJ-type mixed anomaly with the gauged $\IZ_2^F$ fermion number symmetry, so the chiral transformation generates an anomalous shift of the action 
 \eq{
 \Delta S=\pi i\, \CI[a \cdot \rho]~. 
 }
 The partition function of the Majorana fermion with $\IZ_2^F$ is gauged can be written as 
 \eq{
 Z_{Maj/\IZ_2}[B]=\frac{1}{\sqrt{|H^1|}}\sum_{a \in H^1(M_2;\IZ_2)} Z_{Maj}[a ]\times (-1)^{ \int B\cup a}~. 
 }
 The $\IZ_2^c$ transformation acts on the partition function by 
 \eq{
\IZ_2^c: Z_{Maj/\IZ_2}[B]\longmapsto \frac{1}{\sqrt{|H^1|}}\sum_{a \in H^1(M_2;\IZ_2)} Z_{Maj}[a ]\times (-1)^{\int B\cup a+\CI[a\cdot \rho]}~. 
 }
 
 We can turn this ABJ-anomalous symmetry into a true (non-invertible) symmetry by gauging the dual $\IZ_2^{(0)}$ symmetry:
 \eq{
\CS:Z_{Maj/\IZ_2}[B]~\longmapsto ~
\frac{1}{\sqrt{|H^1|}}\sum_b Z_{Maj/\IZ_2}[b]\times (-1)^{\CI[b\cdot \rho]+\CI[B\cdot\rho]+\int b\cup B}~. 
 }
 The $\IZ_2^c$ transformation combined with the $\CS$ action is then a symmetry of the theory:
 \eq{
\CW: Z_{Maj/\IZ_2}[B]&\longmapsto \frac{1}{|H^1|}\sum_{a,b}Z_{Maj}[a] \times(-1)^{\int a\cup b+\int b\cup B+\CI[a\cdot \rho]+\CI[b\cdot \rho]+\CI[B\cdot \rho]}\\
&=Z_{Maj/\IZ_2}[B]
 } 
where we used the orientability of spacetime. Since $B\in H^1(M;\IZ_2)$ is the background gauge field for the $\IZ_2$ symmetry of the Ising model which is generated by the $a_1$ Wilson lines (sometimes labeled as $\eta$-lines in the literature), we can interpret the local  sum over $b$ as the 1-gauging of the $\IZ_2$ symmetry of the Ising model. To construct a topological symmetry operator for this symmetry, we would need to perform the gauging on the half-space defined by the symmetry operator world-line. See \cite{Choi:2022zal,Choi:2021kmx} for more details. 

Consider acting twice with the Kramers-Wannier transformation: 
 \eq{
  \CW^2:Z_{Maj/\IZ_2}[B]&=\frac{1}{|H^1|^{3/2}}\sum_{a,b,c}Z_{Maj}[a] \times(-1)^{\int a\cup b+\int b\cup c+\int c\cup B+\CI[B\cdot \rho]+\CI[b\cdot \rho]}\\
 &=\frac{1}{\sqrt{|H^1|}}\sum_{a,b}Z_{Maj}[a] \times(-1)^{\int a\cup b+\CI[B\cdot \rho]+\CI[b\cdot \rho]}\,\delta(b+B)
 \\ &
 =Z_{Maj/\IZ_2}[B]
 }
Implementing the gauging for $\CW^2$ operation on a half-space, which also describes the insertion of a topological symmetry operator, results in a condensation defect.  This follows from the fact that the constraint imposed by integrating over $c$  is relaxed at the topological boundary where $c\big{|}_{bnd}=0$, as in \cite{Choi:2022zal,Choi:2021kmx}. This is how the $\IZ_2$ condensation defect of the Kramers-Wannier duality arises even though there is no remnant of the non-invertibility of Kramers-Wannier duality on the partition function. 
\vspace{0.25cm}

Deforming away from the CFT via an added mass term 
\eq{
\CL_m=i m \bar\chi\gamma^3\chi~ 
}
explicitly breaks the $\IZ_2$ Kramers-Wannier symmetry, as  
\eq{
\IZ_2^c:\bar\chi\gamma^3\chi\longmapsto -\bar \chi\gamma^3\chi~. 
}
The broken Kramers-Wannier symmetry then leads to a non-invertible family of theories 
\eq{
\CW:Z_{Maj/\IZ_2}[m;B]\longmapsto Z_{Maj/\IZ_2}[-m;B]~,
}
where $\CW$ is the composition of a discrete gauging and $\IZ_2^c$ transformation. This is just a rephrasing of the classic Kramers-Wannier duality of the $2d$ Ising model.

\subsection{Generalized Families from Compactification}

\label{sec:compactification}

Another way that generalized families of theories emerge is by considering twisted compactifications of theories with higher group or non-invertible global symmetries.   Dimensionally reducing generalized symmetries was discussed in e.g. \cite{Gukov:2020btk,Nardoni:2024sos,Kaidi:2022uux,Giacomelli:2024sex,Ma:2024hpn,Sheckler:2025fql,Gukov:2025dol}. 

We will first demonstrate how a $d$-dimensional theory with a $U(1)^{(m)}$ global symmetry that participates in an $n$-group for $n>m$ will lead to a higher family of theories when we perform the twisted compactification on an $(m+1)$-manifold.  Consider a $d$-dimensional theory with an $n$-group $\IG_n$ that has a $U(1)^{(m)}$ component where the transformation properties of the component background gauge fields obey\footnote{The background gauge fields of a higher group can have transformations that depend non-linearly on $\lambda_q$, as in e.g.  \cite{Green:1984sg,Cordova:2018cvg,Brennan:2020ehu,Brennan:2023kpw,Benini:2018reh,Hidaka:2020iaz,Hidaka:2020izy,Cordova:2020tij}. We will here restrict our attention to linear transformations for simplicity. }
\eq{
\delta A_{p+1}&=d\lambda_p+\sum_{q<p}\lambda_q\wedge \omega^{(p)}_{p-q+1}(A_{\leq q+1})~,
}
and $A_{\leq q+1}$ are the $\IG_n$ background gauge fields of degree less than or equal to $(q+1)$. We now compactify the theory onto $S^{m+1}$, fixing the holonomy of the degree $(m+1)$-gauge field 
\eq{
\oint_{S^{m+1}}A_{m+1}=\theta~. 
}
Upon compactification the gauge fields $A_p$ for $p>m+1$ generally decompose as 
\eq{
A_{p+1}\longmapsto A_{p+1}+\tildeA_{p-m}\wedge \Omega_{S^{m+1}}+\dots ~, 
}
where $\Omega_{S^{m+1}}$ is the volume form on $S^{m+1}$, and $\dots$ are possible terms from other cycles. 
The large gauge transformation $\delta A_{m+1}=d\lambda_m$ will result in a shift $\theta\mapsto \theta+2\pi$. Because of the $n$-group structure, shifting $\theta\mapsto \theta+2\pi$ is only an identification of the compactified theory if we also shift 
\eq{
 \tildeA_{p-m}\underset{\Delta\theta=2\pi}{\longmapsto} \tildeA_{p-m}+2\pi \,\widetilde\omega_{p-m}^{(p)}(A_{\leq q+1})~,
}
where we have rewritten the higher group variation as
\eq{
\delta A_{p+1}=d\lambda_p+\sum_{q<p}d\lambda_q\wedge \widetilde\omega^{(p)}_{p-q}(A_{<p})~.
}

When we compactify, the holonomies become non-compact parameters in the IR effective theory at energies below the KK scale $1/R_{KK}$. For example, consider a $4d$ theory with Weyl fermions which transform under a $U(1)^{(0)}$ global symmetry. A twisted $S^1$ compactification for this $U(1)^{(0)}$ global symmetry turns on  a $3d$ mass term (whose sign is correlated with the charge of the fermion) which is a non-compact coupling in the low energy theory. 

We can illustrate this decompactification of the holonomy in the example of a single massless $4d$ Weyl fermion that transforms under a $U(1)^{(0)}$ global symmetry with charge $q$. Consider the twisted compactification on $S^1$ of radius $R_{KK}$ which is labeled by the periodic parameter $\theta\in U(1)^{(0)}$. When $\theta$ winds by $\theta\sim \theta+2\pi$, the twisted boundary conditions of the charge $q$ fermion go through $q$ values of $\theta$ that are periodic. These periodic points in $\theta\in S^1_\theta$ correspond to values where the fermion is massless in the low energy $3d$ effective theory. At generic $\theta$ the fermions have mass of order $m\sim 1/R_{KK}$. In order to get low mass fermions, we need to expand near a value  $\theta=\theta_0$ which has massless fermions and take the scaling limit $\theta=\theta_0+\delta \theta$ where $\frac{\delta\theta}{R_{KK}}= m_\psi$ is fixed.  
Thus in the IR limit, the different minima are separated by a potential with height set by the cutoff scale ($E_{\rm cutoff}\sim 1/R_{KK}$) and 
we are effectively limited to a single vacua where $m_\psi$ is non-compact. See for example \cite{Nardoni:2024sos} for further discussion. 

Although the generic twist parameter $\theta$ effectively decompactifies, the higher group symmetry of the UV theory still leads to a higher family of theories in the compactified theory: here there is a higher family of theories with a non-compact parameter that is identified at $\theta\to \pm \infty$ similar to the masses of chiral fermions as studied in \cite{Cordova:2019jnf,Choi:2022odr}. 

\vspace{0.2cm}

To illustrate the inherited family structure, consider the example from Section \ref{sec:highergroup} of $4d$ QED with two Weyl fermions of charge $+1$ and two Weyl fermions of charge $-1$ which transform under a 0-form $U(1)_f\subset SU(2)_+\times SU(2)_-$ 0 global symmetry as 

\begin{center}
\begin{tabular}{c|cc}
&$U(1)_g$&$U(1)_f$\\
\hline 
$\psi_+$&1&$p$\\
$\psi_-$&$-1$&$q$\\
$\chi_+$&1&$-p$\\
$\chi_-$&$-1$&$-q$
\end{tabular}
\end{center}

\noindent  The 2-group symmetry acts on the $U(1)^{(0)}_f$ and $U(1)^{(1)}_m$ background gauge fields as
\eq{
\delta A_f=d\lambda\quad, \quad \delta B_2^{(m)}=d\Lambda_1+\frac{p^2-q^2}{2\pi}\lambda \,dA_f~.
}
with Postnikov class $\kappa\in H^3(BU(1)^{(0)};U(1)^{(1)})$ given by the anomaly coefficient $\kappa=2p^2-2q^2$.  We can now compactify this theory on $S^1$ with twisted boundary conditions for the $U(1)_f$ background gauge field 
\eq{
\oint_{S^1}A_f=\theta~. 
}
In the scaling limit $\theta\to 0$, with $\theta/R_{KK}=m$ fixed, the 3d IR theory will have a $U(1)$ gauge field, a real scalar field, and four fermions with masses:
\eq{
\CL_{mass}=p m\bar\psi_+\psi_++  q m \bar\psi_-\psi_-- pm \bar\chi_+\chi_+- q m \bar\chi_-\chi_-~.
}
If we take $m\to \pm \infty$, we can integrate out the fermions to end up with Maxwell theory with Chern-Simons couplings $S_{\pm \infty}$ that depend on the $3d$ $U(1)_f^{(0)}$ background gauge field:
\eq{
 S_{\pm \infty}=\pm
\frac{ik_{fg}}{2\pi}\int A_f\wedge dA_g\quad, \quad
k_{fg}=2|p|-2|q|~. 
}
If $\kappa \neq 0$ then also $k_{fg}\neq 0$ and there is a higher family of theories.   

In this $3d$ theory, the higher family of theories is expressed by the fact that the shift of the mixed $U(1)_g\times U(1)_f$ Chern-Simons which arises in taking $m\gg0\to m\ll0$ can be canceled by a shift in the background gauge field for the 0-form $U(1)^{(0)}_m$ magnetic symmetry: 
\eq{
S=...+\itwopi\int B_1\wedge dA_g\quad, \quad B_1\longmapsto B_1+2k_{fg} A_f~. 
}
Thus, we find that the partition function obeys
\eq{
Z_{m\to \infty}[B_1]=Z_{m\to -\infty}[B_1+2k_{fg}A_f]~. 
}
The UV constraints of such a higher family structure imply that any $3d$ UV completion (such as a non-abelian gauge theory) that breaks either the $U(1)^{(0)}_f$ or  $U(1)^{(0)}_m$ global symmetries must also break the higher family structure. 

\vspace{0.2cm}
 Similarly, we can consider the twisted compactification of a theory with a non-invertible global symmetry.  This need not lead to a non-invertible family of theories, although it does in special cases.   As an example with  $4d$ non-invertible chiral symmetry which does not lead to a non-invertible family, consider $4d$ $PSU(N)$ gauge theory with a single Weyl fermion in the adjoint representation, i.e. ${\cal N}=1$ SYM.  The $U(1)^{(0)}_R$ chiral symmetry is broken by an ABJ anomaly to a $\IZ_N$ subgroup.  We can still consider the twisted compactification on $S^1$ of the $U(1)^{(0)}\supset \IZ_N$ connection. As in our above discussion of twisted compactification of higher group global symmetries, the holonomy of $U(1)^{(0)}$ will lead to a mass term for the low energy effective $3d$ theory: 
 \eq{
\CL_{mass}= m \bar\psi\psi~.
 }
Now we see that the action differs in the limit $m\to \pm\infty$ by an integral Chern-Simons term that only depends on the dynamical gauge field. Unlike the case of higher group global symmetries, this cannot be canceled by a discrete gauging. Here, the reason that we do not get a non-invertible family structure is that although the integral part of the $PSU(N)$ Chern character (i.e. instanton density) is trivial for a closed $4d$ manifold, its trivialization is a $3d$ Chern-Simons term which is not trivial and does not factorize -- it is generically $U(1)$-valued.  This prevents us from using a topological gauging that realizes the non-invertible family structure. 

This is different than the case of higher group global symmetries because in that case the twisted compactification does not require trivializing any dynamical gauge fields.  
Indeed, this trivialization of the non-invertible structure is avoided in the example of $4d$ QED with a Dirac fermion of charge $+1$.   
This theory has the non-invertible $\IQ/\IZ$ global symmetry and  the twisted compactification yields a correlated mass between the two $3d$ fermions; the limits $m\to \pm\infty$ differ by an integral, abelian Chern-Simons term. As discussed in \cite{Witten:2003ya}, abelian Chern-Simons terms can be shifted by twisted gauging (in fact there is an $SL(2;\IZ)$ action on the theory) and consequently the partition functions $Z_{m\to \infty}$ and $Z_{m\to -\infty}$ can be related if we supplement with an additional gauging, resulting in a non-invertible family structure.

\section{Generalized Family Anomalies}

\label{sec:anomalies}

The notions and constraints of anomalies extend to include~ \emph{family anomalies} \cite{Cordova:2019jnf,Cordova:2019uob,Debray:2023ior,Brennan:2024tlw,Hsin:2020cgg,Kobayashi:2023ajk,Choi:2022odr,Sharon:2020doo,Manjunath:2024rxe}. 
Consider the partition function of a given family of $d$-dimensional theories $\CT_\theta$ on $M_d$ with group-like symmetry $\CG$ which is indexed by a continuous parameter $\theta$. We will write the partition function as $Z_\theta[A]$ where $A$ are the background gauge fields for $\CG$. 
As we have discussed, the theory may be identical at different theta: $\CT_\theta\cong \CT_{\theta+2\pi}$ such as when $\theta\in \CC_\theta=S^1_\theta$. The partition function of the theory may then be compared at these different values of $\theta$. The partition function only needs to be equivalent up to a phase
\eq{\label{standardanomalousphase}
Z_\theta[A]=Z_{\theta+2\pi }[A]\times e^{ i \int  \omega_d(A)}~,
}
where $\omega_d(A)$ is only dependent on background gauge fields $A$. This phase can be matched by an $(d+1)$-dimensional SPT phase with path integrand $e^{i\CA}$ with: 
\eq{\label{standardSPT}
\CA=\int_{Y_{d+1}} \frac{d\theta(y)}{2\pi}\wedge \omega_d(A)~,
}
where $\partial Y_{d+1}=X_d$ and $\theta(y)$ is a function on $Y_{d+1}$ with boundary value $\theta(y)|_{x\in X_d}=\theta$.\footnote{This SPT phase is only meaningful with the identification $\CT_\theta\cong \CT_{\theta+2\pi}$. The SPT phase implicitly comes with such an identification so that it reproduces the relative phase of the partition function. 
} 
Since this phase is topological, the anomaly is matched along RG flows.

This behavior is nicely exemplified in the case of $4d$ $SU(N)$ Yang-Mills, which has a mixed anomaly between the 1-form $\IZ_N^{(1)}$ center symmetry and $\theta \sim \theta +2\pi$ periodicity~\cite{Cordova:2019uob}.  This is seen from the fact that coupling $\IZ_N^{(1)}$ to a background gauge field $B_2\in H^2(X_d;\IZ_N)$ leads to fractional  instanton number \cite{Witten:2000nv,Aharony:2013hda}:
\eq{\label{SUNFamilyAnomaly}
\int \frac{\Tr[F\wedge F]}{8\pi^2}=\frac{N-1}{2N}\int \CP(B_2)~{\rm mod}_\IZ\quad \Longrightarrow\quad \CA=\frac{N-1}{2N}\int d\theta\cup \CP(B_2)~.
}
The action does not obey the periodicity $\theta\sim\theta+2\pi$  in the presence of background $B_2$ gauge fields  
and the partition function behaves as
\eq{
Z_\theta[B_2]=Z_{\theta+2\pi}[B_2]\times e^{-\frac{\pi i (N-1)}{N}\int \CP(B_2)}~.
}
The partition function is a non-trivial line bundle~\cite{Cordova:2019uob} over the space $S^1_\theta\times B^2_\nabla \IZ_N$ where here $B^2_\nabla \IZ_N$ is the space of 2-form $\IZ_N$ gauge fields. 

Since the additional phase above is quantized and only depends on background gauge fields, the behavior cannot be canceled by any gauge invariant $d$-dimensional counter terms and is physically meaningful (i.e. independent of regularization scheme). Due to the similarity with `t Hooft anomalies, which can be interpreted as the partition function being a non-trivial line bundle over the space of background gauge fields, 
this behavior is referred to as an anomaly in the space of coupling constants or a family anomaly \cite{Cordova:2019jnf,Cordova:2019uob,Debray:2023ior,Lichtman:2020nuw,Brennan:2024tlw,Artymowicz:2023erv,Hsin:2020cgg}. In the example above, there is a family anomaly between $\theta$ and $\IZ_N^{(1)}$ global symmetry.

Family anomalies also have physical consequences. Since the family anomaly can be described by a $(d+1)$-dimensional SPT phase, the anomalous behavior must be matched along symmetry preserving RG flows. As discussed in \cite{Cordova:2019jnf,Hsin:2020cgg,Brennan:2024tlw}, the anomaly can be matched by a family of gapped theories with phase transitions or by spontaneously breaking some of the global symmetries that participate in the family anomaly.

If one considers allowing the parameter $\theta$ to vary in $X_d$ as $\theta(x)$, the  $\theta(x)$-configuration that winds from $\theta\mapsto \theta+2\pi $ has an associated world volume anomaly.  For the case of $4d$ $SU(N)$ Yang-Mills, 
allowing $\theta$ to be spacetime dependent (such as in a background axion field), the domain wall configuration indeed carries world-volume 't Hooft anomaly~\cite{Callan:1984sa,Hsin:2018vcg}:
\eq{
\CA_{w.v.}=\frac{\pi  (N-1)}{N}\int \CP(B_2)~. 
}
This can be matched e.g. if the $\IZ_N^{(1)}$ global symmetry is spontaneously broken on the domain wall and the $SU(N)$ gauge theory deconfines there. Also, as $\theta\to \theta+2\pi$, it sweeps through $\theta=\pi$ where there is the mixed anomaly of~\cite{Gaiotto:2017yup} between  $\IZ_N^{(1)}$ and time-reversal, which is consistent with the anomaly of \cite{Cordova:2019jnf,Cordova:2019uob}.

\subsection{Anomalies of Higher Families}

Families of QFTs with higher symmetry acting on the space of coupling constants can likewise have anomalies.  A higher family anomaly is encoded by the partition function being a non-trivial section over the reduced parameter space $\CB/\Lambda=\frac{\CC_\theta\times B_\nabla \CG}{\Lambda}$ as in \eqref{highergrouplinebundle}. Assume that the total group-like symmetry is $\CG=\IG\times G$, where $\IG$ is the higher-group that participates in the higher family structure and $G$ is some other group-like global symmetry that participates only in the family anomaly. There exists a scheme in which the partition function will then only be invariant modulo an anomalous phase under higher family relations that shifts the coupling constant $\theta\mapsto \theta+2\pi$ and the $\IG$ background gauge field $B\mapsto B+\Lambda$:
\eq{
Z_\theta[B]=Z_{\theta+2\pi }[B+\Lambda]\times e^{ i \int \omega_d(A)}~,
}
where $A$ are the background $G$ gauge fields.   The anomaly is matched by inflow via the $(d+1)$-dimensional SPT phase: \eq{\label{intrinsicSPT}
\CA_{fam}=\int \frac{d\theta}{2\pi}\cup \omega_d(A)~,
}
where $\omega_d$ is a quantized $H^d(BG;U(1))$ characteristic class (i.e. $\exp\left\{i \,k\int \omega_d(A)\right\}$ is only $G$-background gauge invariant for $k\in \IZ$) and $\theta$ is the extension of the coupling constant to a periodic function.  As with standard family anomalies, the higher family anomalies are RG invariants that must be matched along RG flows that preserve both the symmetries that participate in the higher family structure ($\IG$) and the symmetries that are involved with the anomaly $(G)$.

As the case with higher group global symmetries, some family anomalies can be canceled by local counter-terms due to the higher family structure. In the case where $\omega_{d}(A)=\Lambda\cup \widetilde\omega(A)$, the anomaly in \eqref{intrinsicSPT} can be canceled by the local counter term:
\eq{
S_{c.t.}= i\int B\cup \widetilde\omega(A)~.
} 
In cases where the counter-term itself is not gauge invariant, the family anomaly may be canceled at the expense of introducing an `t Hooft anomaly for the global symmetry involved in the family structure. The family structure can thus intertwine `t Hooft and family anomalies.

\subsubsection{Example: $4d$ Maxwell Theory}
\label{sec:4dMaxwell}

To illustrate anomalous higher family of theories,  consider $4d$ Maxwell theory with a $\theta$-term: 
\eq{
S_\theta=\int \frac{1}{2g^2}F_2\wedge \ast F_2-\frac{i \theta}{8\pi^2}\int F_2\wedge F_2~. 
}
 Upon coupling the electric and magnetic 1-form global symmetry $U(1)^{(1)}_e\times U(1)^{(1)}_m$ to corresponding background gauge fields $B_2^{(e)},B_2^{(m)}$, the action is
\eq{
S=&\int \frac{1}{2g^2}(F_2-B_2^{(e)})\wedge \ast (F_2-B_2^{(e)})-\frac{i \theta}{8\pi^2}\int (F_2-B_2^{(e)})\wedge (F_2-B_2^{(e)})
\\&
-\itwopi\int B_2^{(m)}\wedge F_2
}
The $U(1)_e^{(1)}\times U(1)^{(1)}_m$ have mixed anomaly~\cite{Gaiotto:2014kfa}
\footnote{As usual with mixed anomalies, we can choose which symmetry to preserve by choice of a local counter term, with the anomaly theory differing by an integration-by-parts surface term.}
\eq{\label{maxwellanomaly}
\CA=\frac{1}{2\pi} \int B_2^{(e)}\wedge dB_2^{(m)}~,
}
which encodes the variation of the partition function 
\eq{\label{MaxwellB2B3Anom}
Z_\theta[B_2^{(e)}+2\pi d\Lambda_1^{(e)},B_2^{(m)}+2\pi d\Lambda _1^{(m)}]=e^{i\int d\Lambda_1^{(e)}\wedge B_2^{(m)}}\, Z_\theta[B_2^{(e)},B_2^{(m)}]~. 
}

The periodicity of $\theta$ is $\theta\sim \theta+4\pi$ on a generic oriented manifold. For a spacetime with spin structure, i.e. if $w_2(TM)=0$, the periodicity is $\theta \sim \theta +2\pi$.  For spacetimes without spin structure (e.g. $M_4=\ICP ^2$), shifting $\theta \sim \theta +2\pi$ can still be regarded as a classical symmetry if combined with a compensating shift of the $B_2^{(m)}$ background: 
\eq{
Z_\theta[0;B_2^{(m)}]=Z_{\theta+2\pi}[0;B_2^{(m)}+\pi w_2(TM)]~. 
}
For non-zero background $B_2^{(e)}$, this higher family structure has an anomaly:
\eq{
Z_\theta[B_2^{(e)},B_2^{(m)}]=Z_{\theta+2\pi}[B_2^{(e)},B_2^{(m)}+\pi w_2(TM)+ B_2^{(e)}]\times e^{-\pi i \int \frac{B_2^{(e)}}{2\pi}\wedge \frac{B_2^{(e)}}{2\pi}}~. 
}
This anomaly of the higher family is described by a 5d anomaly theory 
\eq{\label{maxwellfamilyanomaly}
\CA\supset - \int d\theta \cup \frac{B_2^{(e)}\wedge B_2^{(e)}}{8\pi^2}~.
} 
The full anomaly theory contains both terms \eqref{maxwellanomaly} and  \eqref{maxwellfamilyanomaly}
\eq{
\CA=&\frac{1}{2\pi}\int dB_2^{(m)} \cup B_2^{(e)}- \int d\theta \cup \frac{B_2^{(e)}\wedge B_2^{(e)}}{8\pi^2}~.
}

\subsubsection{Example: $4d$ $SU(4N+2)/\IZ_2$ Yang-Mills}

As another example of an anomalous higher family of theories, consider $SU(4N+2)/\IZ_2$ Yang-Mills, with action:
\eq{
S=\int \frac{1}{2g^2} \Tr[F\wedge \ast F]+\frac{i\theta}{8\pi^2}\int \Tr[F\wedge F]~. 
}
For $SU(2n)/\IZ_2$ gauge theory, the instanton number quantization is
\eq{
\frac{1}{8\pi^2}\int \Tr[F\wedge F]=\frac{n(2n-1)}{4}\int \CP(w_2(\IZ_2))~{\rm mod}_\IZ~,
}
where $w_2(\IZ_2)$ is the characteristic class (dynamically summed over) of the $SU(2n)/\IZ_2$ bundle that obstructs the lift to a $SU(2n)$ bundle. So $SU(4N+2)/\IZ_2$ gauge theory has $\theta\sim \theta+8\pi$ on a generic manifold (prior to turning on background gauge fields  for the 1-form symmetries). 

The theory has a $\IZ_2^{(1)}$ magnetic global symmetry whose background gauge field $B_2^{(m)}$ couples to $w_2(\IZ_2)$ as
\eq{
S\supset \pi i \int w_2(\IZ_2)\cup B_2^{(m)}~. 
}
Under a shift $\theta\mapsto \theta+4\pi$, the action shifts as 
\eq{\label{higherformintegrand} 
S_\theta\mapsto S_{\theta+4\pi}=S_\theta+\pi i \int \CP(w_2(\IZ_2))~. 
}
The second term can be canceled by shifting the background gauge field for the $\IZ_2^{(1)}$ magnetic symmetry\footnote{Via the Wu formula, $\CP(x_2)~{\rm mod}_2=x_2\cup w_2(TM)$ for any $\IZ_2$-valued 2-form $x_2$}, with \eq{
Z_\theta[B_2^{(m)}]=Z_{\theta+4\pi}[B_2^{(m)}+w_2(TM)]~.
}

This theory also has a $\IZ_{2N+1}^{(1)}$ electric global symmetry and turning on its background gauge field $B_2^{(e)}$ affects the quantization of the $SU(2n)/\IZ_2$ instanton number as\footnote{
A general $PSU(2n)$ bundle with obstruction class $b_2$ has fractional instanton number:
\eq{
\frac{1}{8\pi^2}\int \Tr[F\wedge F]=k+\frac{2n-1}{4n}\int \CP(b_2)~.
}
We can then decompose the obstruction class $b_2$ into $\IZ_2$ and $\IZ_n$ parts as $b_2=n w_2(\IZ_2)+B_2^{(e)}$. 
}
\eq{
\int \frac{\Tr[F\wedge F]}{8\pi^2}=&k+\frac{2n-1}{4n}\int \CP(B_2^{(e)})+\frac{2n-1}{2}\int w_2(\IZ_2)\cup B_2^{(e)}\\
&+\frac{n(2n-1)}{4}\int \CP(w_2(\IZ_2))~,
}
for $k\in \IZ$. This implies that for $SU(4N+2)/\IZ_2$ with a background $\IZ_{2N+1}^{(1)}$ gauge field turned on, the shifting $\theta\mapsto \theta+4\pi$ and $B_2^{(m)}\mapsto B_2^{(m)}+w_2(TM)$, leads to a shift of the partition function:
\eq{
Z_\theta[B_2^{(e)},B_2^{(m)}]=Z_{\theta+4\pi}[B_2^{(e)},B_2^{(m)}+w_2(TM)]\times e^{\frac{\pi i}{2N+1}\int \CP(B_2^{(e)})}~.
}
Therefore, this higher family of theories is anomalous:
\eq{
\CA=\int \frac{d\theta}{4(2N+1)}\cup \CP(B_2^{(e)})~.
}
 This example does not have an additional anomaly for the $\IZ_2^{(1)}$ global symmetry.\footnote{There is no mixed anomaly between $\IZ_2^{(1)}$ and $\IZ_{2N+1}^{(1)}$ because the center of $SU(4N+2)$ is a direct product $Z(SU(4N+2))=\IZ_2\times \IZ_{2N+1}$. On the other hand, for $SU(4N)/\IZ_2$ Yang-Mills, the $\IZ_2^{(1)}$ magnetic symmetry does have a mixed anomaly with the $\IZ_{2N}^{(1)}$ center symmetry. In this case the analogous anomaly for the higher family structure will exhibit the higher family structure, as in the case of Maxwell theory in Section \ref{sec:4dMaxwell}.}

\subsection{Anomalies of Categorical Families}

Categorical families of theories can also be anomalous.  Consider a non-invertible family of theories with a $\CG=G_A\times G_B$ global symmetry where $G_A$ participate in the non-invertible family structure and $G_B$ do not. 
As in the case of a higher family of theories, these anomalies can be defined by whether
or not the partition function is a non-trivial line bundle over the reduced parameter space $\CB/\Lambda=\frac{\CC_\theta\times \fB}{\Lambda}$ as in \eqref{noninvertiblelinebundle}.  There is a non-invertible family anomaly if the partition function picks up a non-trivial phase when shifting by the family identification:
\eq{
\CS\cdot Z_{\theta+2\pi}[B]=Z_{\theta}[B]\times e^{i \int \omega_d(B)}~,
}
where $B$ are the gauge fields for the $G_B$ group-like global symmetry.  
The  anomaly can then be characterized by an SPT phase of the form 
\eq{
\CA= \int d\theta \cup \omega_d(B)~.
}
This SPT phase is only meaningful with the non-invertible family identification,
\eq{
\CT_\theta\cong \CS\CT_{\theta+2\pi}~,
}
comparing partition functions of identified theories. This is as with the standard family anomaly: the SPT phase \eqref{standardSPT} is only meaningful with identification $\CT_\theta\cong \CT_{\theta+2\pi}$,  reproducing the anomalous phase of the family relation on the partition function in \eqref{standardanomalousphase}. 
Again, the categorical family anomaly must be matched for all $\CG$-preserving continuous deformations including, in particular, RG flows.

\subsubsection{Example: $SU(2NM)/\IZ_{2M}$ Yang-Mills and $T$-Symmetry}
\label{sec:masterEx}

As an example of a family of QFTs with a categorical structure that includes both a higher family and non-invertible family structures that are both anomalous, consider $4d$ $SU(2NM)/\IZ_{2M}$ Yang-Mills theory with a $\theta$-angle:
\eq{
S=\int \frac{1}{2g^2}\Tr[F\wedge \ast F]+\frac{i \theta}{8\pi^2}\int \Tr[F\wedge F]~.
}
We assume for simplicity that $N,2M$ are co-prime. 

This theory has a $\IZ_{N}^{(1)}$ center symmetry and $\IZ_{2M}^{(1)}$ magnetic symmetry which do not have a mixed anomaly since $N,2M$ are co-prime. 
We can turn on a background gauge field $B_2^{(e)}$ for the $\IZ_N^{(1)}$ by restricting the path integral to $PSU(2NM)$ bundles with fixed obstruction class mod$_{N}$.  The full $\IZ_{2NM}$-valued obstruction class $w_2(\IZ_{2NM})$  can be decomposed as
\eq{
w_2(\IZ_{2NM})=N w_2(\IZ_{2M})+2M B_2^{(e)}~, 
}
where $w_2(\IZ_{2M})$ is the obstruction class for the dynamical $SU(2NM)/\IZ_{2M}$-bundle to be lifted to a $SU(2NM)$-bundle. 
With respect to this decomposition, the instanton number is fractional
\eq{
\int \frac{\Tr[F\wedge F]}{8\pi^2} 
=\frac{M(2NM-1)}{N}\int \CP(B_2^{(e)})+\frac{N(2NM-1)}{4M}\int \CP(w_2(\IZ_{2M}))~{\rm mod}_\IZ~.  
}

First setting $B_2^{(e)}=0$,  the evident periodicity $\theta\sim \theta+8\pi M$ can be enhanced to $\theta\sim \theta+4\pi M$ by introducing an additional shift of the 1-form magnetic symmetry $B_2^{(m)}$:
\eq{
\Delta\theta=4\pi M\quad, \quad B_2^{(m)}\underset{~}{\longmapsto} B_2^{(m)}+Mw_2(TM)
~,
}
where $B_2^{(m)}$ couples as 
\eq{
S=...+
\frac{2\pi i}{2M}\int B_2^{(m)}\cup w_2(\IZ_{2M})~.
}
For non-zero $B_2^{(e)}$, the partition function transforms as
\eq{
Z_\theta[B_2^{(e)};B_2^{(m)}]=Z_{\theta+4\pi M}[B_2^{(e)};B_2^{(m)}+Mw_2(TM)]\times e^{\frac{4\pi i M^2}{N}\int \CP(B_2^{(e)})}~,
}
which indicates that the higher family has an inherent anomaly:
\eq{\label{superexampleanomSPT}
\CA=-\frac{M}{N}\int d\theta\cup \CP(B_2^{(e)})~. 
}

This family of theories also has a non-invertible structure which relates $\theta\sim \theta+2\pi$ if we combine it with the discrete gauging operation $\CS$ which changes the action
\eq{
\CS\cdot S[w_2(\IZ_{2M})]\mapsto S[w_2(\IZ_{2M})]+i \int w_2(\IZ_{2M})\cup b_2+\frac{2iM}{2\pi}\int b_2\cup dc_1+\frac{i M}{2\pi}b_2\cup b_2~,
}
where $b_2,c_1$ are dynamical $U(1)$ gauge fields (summed over in the path integral). 

To focus on the non-invertible family structure, let us restrict our theory to spin manifolds. 
When $B_2^{(e)}\neq 0$, the partition function transforms as 
\eq{
\CS\cdot Z_{\theta}\left[B_2^{(e)},B_2^{(m)}\right]=Z_{\theta+2\pi}\left[B_2^{(e)},B_2^{(m)}\right]\times e^{\frac{2\pi i M}{N}\int \CP(B_2^{(e)})}~. 
}
This anomaly is also captured by the SPT phase \eqref{superexampleanomSPT} due to the different shift of $\theta$.  

This example also has an interesting interplay with time reversal symmetry. As described in \cite{Choi:2022rfe}, $SU(2NM)/\IZ_{2M}$ Yang-Mills has a non-invertible time reversal symmetry at $\theta= 
\pi n$ for $n\in \IZ$. This non-invertible time reversal symmetry was constructed by noting that at $\theta= 
\pi n $, time reversal symmetry maps $T:\theta\mapsto -\theta$ and the effective shift $\Delta \theta=2\pi n$ can be undone by the $\CS$ gauging operation so that the combined $\CS\cdot T$ is a symmetry of the partition function (with $B_2^{(e)}=0$). Since the combined action of $\CS\cdot T$ is non-invertible, it can be interpreted as a non-invertible symmetry. 

Here, our discussion suggests a complementary viewpoint where we think of $T$ as having a group-like action, but then we realize that the identification $\theta\sim \theta+2\pi n$ together with the $\CS$ action is part of the categorical structure of the family of theories. This is the non-invertible analog of how the $T\times \IZ_N^{(1)}$ mixed anomaly of $4d$ $SU(N)$ Yang-Mills \cite{Gaiotto:2017yup} also comes from the anomaly described in \eqref{SUNFamilyAnomaly}.

\subsubsection{Categorical Families of QFTs from Gauging}
\label{sec:gauging}

Now we will describe a general mechanism for generating non-invertible families of theories from anomalous continuous families of theories.  
 Consider a family of theories with $G$ symmetry which is indexed by a single parameter $\theta\in S^1$, and suppose that the family has an anomaly of the form 
\eq{
\CA= \int d\theta\wedge \omega_d(A)~,
}
where $A$ are the background $G$ gauge fields   and $\theta$ has period $\theta\sim \theta+2\pi$. 

If we gauge a subgroup $H\subseteq G$, then we can decompose the anomaly into three terms:
\eq{
\CA= \int d\theta\wedge \left(\omega_d(a)^{(0)}+\omega_d^{(1)}(a,A)+\omega_d^{(2)}(A)\right)~,
}
where $a$ are the dynamical gauge fields for $H$ and $A$ are the background gauge fields for $G/H$. The term $\omega_d^{(0)}$ is only dependent on the dynamical gauge field and constitutes a sort of ``family ABJ anomaly'' which one would be tempted to say breaks some of the family identifications. 
The term $\omega_d^{(2)}$ is only dependent on the background gauge field and corresponds to the remaining family anomaly. The term $\omega_d^{(1)}$ is dependent on both $A$ and $a$ which we assume vanishes for $A=0$.

When $H$ is an abelian group, $\omega_d^{(0)}$ factorizes into a polynomial in $a$ and its field strength (or Bockstein as appropriate). In this case the phase can often be eliminated by a non-invertible gauging (with torsion) of the dual, emergent symmetry which generalizes the $\CS_q$ operation described in the previous section. This allows us to retain the periodicity $\theta\sim \theta+2\pi$ and results in a non-invertible family of QFTs. 

Similarly, $\omega_d^{(1)}$ can be further decomposed into terms which are a polynomial in $a,\delta a$ times a lower-dimensional quantized phase:
\eq{
\omega_d^{(1)}=\sum_n p_n(a,\delta a)\wedge \Omega^{(d)}_n(A)~, 
}
where the $\Omega^{(d)}_n(A)$ only depend on the background gauge fields. In this decomposition each $p_n(a,\delta a)$ must be quantized (since $\omega_d^{(1)}$ is quantized) and therefore can be coupled to a background gauge field $C^{(n)}$. The background gauge transformations of the $C^{(n)}$ can then be modified so that the terms in $\omega^{(1)}_d$ that are generated by the shift $\theta\mapsto \theta+2\pi$ are removed. This results in a higher family structure. 

The categorical family structure that these three terms generate mimics the categorical symmetry structure that arises from gauging anomalous symmetries. 
\vspace{0.25cm}

An illustrative example of this behavior is $4d$ $SU(NM)$ Yang-Mills theory where $N,M$ are co-prime. In this theory, there is a $\IZ_{NM}^{(1)}$ center symmetry. As discussed earlier, this describes a family of theories by turning on a $\theta$ term 
\eq{
S=...+\frac{i \theta}{8\pi^2}\int \Tr[F\wedge F]~,
}
where $\theta\sim \theta+2\pi$. This family of theories is anomalous with anomaly
\eq{
\CA=\frac{NM-1}{2NM}\int d\theta\cup \CP(B_2^{(e)})~, 
}
where $B_2^{(e)}$ is the background $\IZ_{NM}^{(1)}$ gauge field. 

Now consider gauging $\IZ_M\subset \IZ_{NM}^{(1)}$. If we decompose 
\eq{
B_2^{(e)}=N w_2(\IZ_M)+M \tildeB_2^{(e)}~,
}
where $w_2(\IZ_M)$ are summed over in the path integral and $\tildeB_2^{(e)}$ is the background gauge field for the remaining $\IZ_N^{(1)}$ center symmetry. As discussed above, the anomaly now decomposes into two terms:
\eq{
\CA=\frac{N(NM-1)}{2M}\int d\theta\cup \CP(w_2)+\frac{M(NM-1)}{2N}\int d\theta\cup \CP(\tildeB_2^{(e)})~. 
}
The first term can be interpreted as an ABJ anomaly for the family whereas the second is a standard family `t Hooft anomaly and  implies that the integrand of the partition function transforms as 
\eq{
\sum_{w_2}\int [da]~Z_{\theta+2\pi}[a;\tildeB_2^{(e)}]=
\sum_{w_2}\int [da]~Z_\theta[a;\tildeB_2^{(e)}]\times e^{\pi i (NM-1)\left( \frac{N}{M}\int \CP(w_2)+\frac{M}{N}\int \CP(\tildeB_2^{(e)})\right)}
}
thereby breaking the standard family relation.  The additional $w_2$-dependent phase in the integrand can be canceled by a discrete gauging which leads to a non-invertible family structure. The remaining $\tildeB_2^{(e)}$-dependent phase can then be interpreted as an anomaly for this new, non-invertible family structure.

\subsection{Domain Walls of Generalized Theories}

Another feature of families of theories indexed by a continuous parameter is that the family structure admits additional defect operators. We  first consider the example of a family of theories parametrized by $\theta\in S^1$. 
A defect for such a family is the interface that interpolates between different theories of the same family. These can be constructed by allowing the parameter $\theta$ to jump discontinuously across a co-dimension 1 manifold $\Sigma$. These defects can be constructed as a singular limit of continuous configurations in where $\theta$ is a spacetime dependent function $\theta(x)$ which smoothly winds through $S^1$ \cite{Hsin:2018vcg,Cordova:2019jnf,Cordova:2019uob}.\footnote{For example, we can take $\theta(x)$ to have a profile which is controlled by a parameter $\epsilon$ such that when we take $\epsilon\to 0$, the spacetime dependent parameter $\theta(x)\to \Omega(x\in \Sigma)$ which is the global angular form for $\Sigma$.} 

When the family of theories is anomalous, these defects have additional physical consequences. 
Consider a continuous family of $\CG$-symmetric theories with anomaly 
\eq{\label{ordinaryfamilyanomaly}
\CA= \int d\theta \cup \omega_d(A)~,
}
where $\omega_d(A)$ is dependent on the $\CG$ background gauge fields. Now consider the interface where $\theta\mapsto \theta+2\pi$. This interface is a defect operator in the theory $\CT_\theta$ since it is an interface between the theory and itself $\CT_{\theta+2\pi}\cong \CT_\theta$ by using the family relation. Because $\theta$ jumps discontinuously across the domain wall, the defect operator will activate the anomaly, resulting in a world-volume anomaly on the defect 
\eq{\label{famworldvolumeanom}
\CA_{w.v.}=2\pi  \int \omega_d(A)~. 
}
This implies that there are non-trivial quantum degrees of freedom living on the domain wall which, depending on the type of anomaly in \eqref{famworldvolumeanom}, may imply that the world volume theory is gapless  \cite{Lieb:1961fr,Wang:2014lca,Wang:2016gqj,Wang:2017loc,Kobayashi:2018yuk,Wan:2018djl,Wan:2019oyr,Cordova:2019bsd,Cordova:2019jqi,Apte:2022xtu,Brennan:2023kpo,Brennan:2023ynm,Brennan:2024tlw,Hsin:2020cgg,Damia:2023ses,Antinucci:2023ezl}.

\vspace{0.25cm}
As in the case of ordinary families of theories, generalized families of theories can also have interesting domain wall structures.  We first consider the case of higher families of theories where the family structure involves a $\IG_n$ $n$-group global symmetry with background gauge fields $B$. If we take the higher family relation to be  $\theta\sim \theta+2\pi$ and $B\mapsto B+\Lambda_\IG$, the interface where $\theta\mapsto \theta+2\pi$ is not a domain wall in the theory $\CT_\theta$.  If we dress the interface by a collection of symmetry defect operators for $\IG_n$ which implement the shift $\Lambda_\IG$ of the background gauge fields $B$, then the interface will become a defect in the theory $\CT_\theta$ due to the higher family structure. If the higher family structure is anomalous, the domain wall will again carry a world volume anomaly which, as in the case of ordinary family anomalies, will require a dynamical quantum theory on the defect that matches the anomaly. 
In particular, for higher family anomalies involving $\IG_n$ of the form in \eqref{ordinaryfamilyanomaly} the higher family domain wall will carry the anomaly given in \eqref{famworldvolumeanom}.  

For non-invertible families of theories that arise from gauging a symmetry $\CG$ so that $\CT_\theta\cong \CS\CT_{\theta+2\pi q}$ for some $q\in \IQ/\IZ$, 
 the interface where $\theta\mapsto \theta+2\pi q$ is also not a defect in the theory $\CT_\theta$.  However, if we modify the interface by condensing a collection of $\CG$ topological symmetry operators (with a twist if necessary) which locally gauges $\CG$, then we can elevate the interface to a defect operator due to the identification $\CT_\theta\cong \CS\CT_{\theta+2\pi q}$. Again, if the non-invertible family structure is anomalous, this domain wall will have a world-volume anomaly. As in the case of ordinary family anomalies, this will imply that there are quantum degrees of freedom living on the domain wall and the anomaly will constrain the possible defect theory matching the anomaly.

\subsection{Constraints on IR Phases from Anomalies of Generalized Families}

We now consider how the anomalies of generalized families of QFTs constrain the possible IR phase  structure of the family of theories. First, we will review the constraints from an anomalous family of theories without any generalized family structure \cite{Brennan:2024tlw}. 

Consider a family of $d$-dimensional theories 
$\CT_\theta$ parametrized by a continuous parameter $\theta\in S^1$ where $\CT_\theta\cong \CT_{\theta+2\pi}$ with a discrete global symmetry group $\CG=G^{(p-1)}\times G^{(d-p-1)}$ that carries a family anomaly of the form 
\eq{
\CA=\frac{2\pi  }{N}\int \frac{d\theta}{2\pi}\cup A_p\cup B_{d-p}~,
}
where $A_p,B_{d-p}$ are background gauge fields for $G^{(p-1)},G^{(d-p-1)}$ respectively. We assume that $\CG$ does not participate in any other anomalies so that the partition function obeys 
\eq{
Z_\theta[M_d;A_p+d\lambda_{p-1},B_{d-p}+d\Lambda_{d-p-1}]=Z_\theta[M_d;A_p,B_{d-p}]\quad, \quad \forall \theta\in S^1~.
}
The family anomaly implies that the partition function obeys the relation
\eq{\label{OrdinaryFamAnomalousPhase}
Z_\theta[M_d;A_p,B_{d-p}]=Z_{\theta+2\pi}[M_d;A_p,B_{d-p}]\times e^{-\frac{2\pi i}{N}\int A_p\cup B_{d-p}}~.
} 
Let us determine the conditions for when this family of theories can flow to a family of gapped phases. As proven in \cite{Cordova:2019bsd}, if the family of theories flows to a family of unitary, $\CG$-preserving gapped phases -- i.e. $\CG$ is not spontaneously broken at any $\theta$ -- then the partition function is non-zero on $S^p\times S^{d-p}$:
\eq{
Z_\theta[S^p\times S^{d-p};A_p,B_{d-p}]\neq 0\quad, \quad \forall \theta\in S^1~. 
}
If we were to assume what we want to disprove, that the partition function is a smooth function of $\theta$ without phase transitions, we can define the continuous function of $\theta$:
\eq{
e^{i \Phi(\theta,A,B)}=\frac{Z_\theta[S^p\times S^{d-p};A_p,B_{d-p}]}{\big|Z_\theta[S^p\times S^{d-p};A_p,B_{d-p}]\big|}~.
}
As in \cite{Brennan:2024tlw} we can use the anomalous phase relation of the path integral \eqref{OrdinaryFamAnomalousPhase} to derive a contradiction. If we choose $A_p,B_{d-p}$ so that the phase is non-trivial on $S^p\times S^{d-p}$, the function $\Phi$ must be of the form 
\eq{
\Phi(\theta,A,B)=f(\theta)\int A_p\cup B_{d-p}+\Theta(\theta,A,B)~,
}
where the variation of the partition function \eqref{OrdinaryFamAnomalousPhase} implies that 
\eq{
f(\theta+2\pi)=f(\theta)+\frac{2\pi}{N}\quad, \quad \Theta(\theta+2\pi,A,B)=\Theta(\theta,A,B)~.
}
Gauge invariance implies that the functions $f(\theta),\Theta(\theta,A,B)$ obey 
\eq{
f(\theta)\in \frac{2\pi}{N}\IZ~,~\forall \theta\in S^1\quad, \quad \Theta(\theta,A+d\Lambda,B+d\Lambda')=\Theta(\theta,A,B)~,}
The term proportional to $f(\theta)$ is necessary to match the anomaly.   But, since no continuous functions $f(\theta)$ exist that obey both of these conditions, the continuous phase $\Phi(\theta,A,B)$ cannot reproduce the anomaly and the partition function of a continuous family of unitary, gapped, $\CG$-preserving phases cannot carry the family anomaly. This implies that the IR theory must be either 1.) gapless, 2.) spontaneously break $\CG$, or 3.) exhibit a phase transition (i.e. the partition function is discontinuous in $\theta$).

\subsubsection{IR Constraints from Higher Family Anomalies}
We now consider the case of a $d$-dimensional higher family of theories with finite $\CG=G^{(p-1)}\times G^{(d-p-1)}\times G^{(q-1)}$ global symmetry.  We denote the background gauge fields for the different components of $\CG$ as $A_p,~A_{d-p}$, and $B_{q}$ respectively where 
$G^{(q-1)}$ participates in the higher family structure:
\eq{\label{higherfamstructure}
\theta\sim \theta+2\pi \quad, \quad B_q\sim B_q+\omega_q(A) ~,
}
where $\omega_q(A)$ is a $q$-form characteristic class dependent on $A_p,A_{d-p}$ and possible discrete gravitational terms (or additional variation which does not participate in the anomaly/shift under the higher family identification).  
Here we will consider the case where the higher family structure has an anomaly of the form 
\eq{\label{higherfamanomaly}
\CA=\frac{2\pi  K}{N}\int \frac{d\theta}{2\pi}\cup A_p\cup A_{d-p}~,
}
where $K\in \IZ$ and $\CG$ does not participate in any other anomalies. If $\CG$ is preserved, then the partition function obeys 
\begin{align}
&Z_\theta[M_d;A_p,A_{d-p},B_q]=Z_\theta[M_d;A_p+d\lambda_{p-1},A_{d-p}+d\lambda_{d-p-1},B_q+d\Lambda_{q-1}]\quad, \quad \forall \theta~
\label{gaugeinvhigherfamanom}
\end{align}
and the family anomaly implies 
\eq{
Z_\theta[M_d;A_p,A_{d-p},B_q]=Z_{\theta+2\pi}[M_d;A_p,A_{d-p},B_q+\omega_q(A) ]  \times e^{\frac{2\pi i K}{N}\int A_p\cup A_{d-p}}~.
}
Now let us derive under what conditions the above anomalous higher family of theories can flow to a trivially gapped phase. Again, the results of \cite{Cordova:2019bsd} imply that the theory on $S^p\times S^{d-p}$ has a non-vanishing partition function if the IR family of theories is a family of unitary, gapped $\CG$-preserving theories:
\eq{
Z_\theta[S^p\times S^{d-p};A_p,A_{d-p},B_q]\neq 0 \quad, \quad \forall \theta~. 
}
Now notice that if $q\neq p, d-p$, then the higher family structure on $S^p\times S^{d-p}$ is trivialized and the higher family behaves as an ordinary family of theories which obey the same constraints on their IR phases from anomalies as above. Therefore, without loss of generality, we can consider the case where $q=p$. In this case, $\omega_q(A)=k A_p$ is the only possibility up to possible gravitational terms. If $K=kL$ mod$_N$, then the anomaly can be canceled by the SPT phase 
\eq{
Z_{\rm SPT}=e^{-\frac{2\pi i L }{N}\int B_p\cup A_{d-p}}~,
}
if there are no gravitational contributions to $\omega_q$. The case where the family anomaly can be matched by the SPT is exactly the case when there is no anomaly, so assume that either $\omega_q$ has a gravitational contribution or $K\neq kL$ mod$_N$ for some $L\in \IZ_N$. 

We again assume that the theory flows to a smooth higher family of $\CG$-preserving gapped phases for all $\theta$, i.e. that the partition function $Z_\theta[M_d;A,B]$ is smooth, non-zero function of $\theta$. We can then define the phase 
\eq{
e^{ i \Phi(\theta,A_p,A_{d-p},B_p)}=\frac{Z_\theta[S^p\times S^{d-p};A_p,A_{d-p},B_p]}{\big|Z_\theta[S^p\times S^{d-p};A_p,A_{d-p},B_p]\big|}~. 
}

As before, we can show that the constraints imposed on the partition function by assuming that the IR theory is a unitarity, gapped, $\CG$-preserving, continuous family of theories cannot accommodate the family anomaly in \eqref{higherfamanomaly}.\footnote{Here we 
use the fact that the bosonic $\CG$-SPT phases have a cohomological classification \cite{Chen:2011pg,Kapustin:2014tfa,Kapustin:2014zva}.} 
The anomaly implies that the phase function must take the form  
\eq{
\Phi(\theta,A_p,A_{d-p},B_p)=&f(\theta)\int A_p\cup A_{d-p}+\frac{2\pi L}{N}\int B_p\cup A_{d-p} 
+\Theta(\theta,A_p,A_{d-p},B_p)~.
}
When $K\neq k L$ mod$_N$, the anomaly can only be matched by requiring that 
\eq{\label{higherfamcontinuity}
f(\theta+2\pi)=f(\theta)+\frac{2\pi K}{L}\quad, \quad L=0~,
}
If $f(\theta)\not\in \frac{2\pi}{N}\IZ$, then the path integral would not be invariant under the background gauge transformations \eqref{gaugeinvhigherfamanom}. There are then no continuous $f(\theta) $ that satisfy both this quantization condition and semi-periodicity in \eqref{higherfamcontinuity}.   

 This implies that if $K\neq kL$ mod$_N$, an anomalous higher family of QFTs must flow in the IR to 1.) a gapless family, 2.) a family which spontaneously breaks $\CG$, or 3.) a discontinuous higher family of theories (i.e. with a phase transition).

\subsubsection{IR Constraints from Non-Invertible Family Anomalies}
\label{sec:NonInvertibleAnomalyConstraints}

Now consider a non-invertible (categorical) family of $d$-dimensional theories with finite abelian  symmetry group $\CG=G^{(p-1)}\times \tildeG^{(p-1)}$ with background gauge fields $A_p,B_p$ respectively where $d=2p$.\footnote{Here we have restricted to the case where $p=d/2$ because non-invertible families that arise from discrete gauging require a symmetry whose dual symmetry has the same degree.} Assume that the family obeys the categorical family relation 
\eq{
\CT_\theta[A_p,B_p]\cong \CS\CT_{\theta+2\pi}[A_p,B_p]~, 
}
where $\CS$ is the gauging of $\tildeG^{(p-1)}$ global symmetry (with torsion) given by: 
\eq{
\CS Z_{\theta}[A,B]&=\frac{1}{\sqrt{|H^p|}}\sum_{b\in H^2(M;\IZ_N)}Z_{\theta}[A,b] \,e^{\frac{2\pi i}{2N}\int 2 R b\cup B+M\CP(b)+L\CP(B)}
}
for coefficients $M,L,R\in \IZ$ such that $M,L\in 2\IZ$ for $N$ odd and $M,L\in \IZ$ for $N$ even and $\CP$ is the Pontryagin square operation (for $N$ odd $\CP(A_p)=A_p\cup A_p\in H^{2p}(M;\IZ_N)$).  The discussion in Section \ref{sec:NonInvertible} primarily relied on the gauging operation $\CS_q$ which had $R=1,L=0,M=k$ for $q=p/N$ and  $pk=1~{\rm mod}_{\gamma(N)N}$. 
Additionally, assume that the categorical family has an  anomaly of the form
\eq{\label{noninvertiblefamanomaly}
\CA=\frac{2\pi K }{2N}\int \frac{d\theta}{2\pi}\cup \CP(A_p)~,
}
where $K\in 2\IZ$ if $N$ is odd, $K \in \IZ$ if $N$ is even. 
The family anomaly implies that the partition function obeys the relation 
\eq{\label{noninvertibleanomphase}
\CS Z_\theta[A_p,B_p]=Z_{\theta+2\pi}[A_p,B_p]\times e^{-\frac{2\pi i K}{2N}\int \CP(A_p)}~. 
}

\vspace{0.25cm}
This family anomaly trivializes if it can be matched by a continuous (non-invertible) family of $\CG$-preserving unitary gapped phases. Again, a gapped, unitary $\CG$-preserving theory implies that 
\eq{
Z_\theta[S^p\times S^p;A_p,B_p]\neq 0\quad, \quad \forall \theta~.
}
If the anomaly is trivial, we can assume that the family flows to a continuous family of theories in the IR, with the partition function a continuous function of $\theta$ for fixed $A_p,B_p$.  We can then  again define the phase 
\eq{\label{noninvertiblePhi} 
e^{i \Phi(\theta,A,B)}=\frac{Z_\theta[S^p\times S^p;A_p,B_p]}{\big|Z_\theta[S^p\times S^p;A_p,B_p]\big|}~. 
}
Here we can parametrize all choices of $\Phi$.  
The only gauge invariant terms that can contribute to $\Phi$ are the characteristic classes:
\eq{\label{noninvphase}
\Phi(\theta,A_p,B_p)=f(\theta) \int \CP(A_p)+\frac{2\pi }{N}\int \left(k_1 A_p\cup B_p+ \frac{k_2}{2} \CP(B_p)\right)+\Theta (\theta,A_p)~,
}
where again $k_2\in 2\IZ$ for $N$ odd and $k_2\in \IZ$ for $N$ even and $\Theta$ is $\theta$-periodic and independent of $B_p$. 
Using similar arguments as before, we can determine that the $f(\theta)\in \frac{2\pi}{2N}\IZ$ so it can be absorbed into the  
 final term in \eqref{noninvphase} which does not participate in the gauging/anomaly. 

As computed in Appendix \ref{app:NonInvertible}, the SPT terms proportional to $k_1,k_2$ can reproduce the anomalous phase only if $L,R,M$ obey the constraint 
\eq{L+2R+M=0~{\rm mod}_{2N}~,
}
and there exists a solution to the equation
\eq{
k_1^2=-R K~{\rm mod}_{\gamma(N) N}~.
} 
In this case the anomaly trivializes: the putative anomalous phase can be cancelled by dressing the theory by the SPT $Z_{SPT}=e^{\frac{2\pi i k_1}{N}\int A_p\cup B_p+ \frac{2\pi i k_2}{2N}\int \CP(B_p)}$. 
A similar discussion appeared in \cite{Apte:2022xtu} on SPT phases with non-invertible global symmetries.   When these trivializing 
conditions hold, there is no genuine anomaly as the putative family anomaly can be canceled by a choice of local counter-terms for a continuous family of SPTs.  

When the anomaly does {\it not} trivialize, the non-invertible family anomaly is a genuine anomaly.  In this genuine anomaly case, a non-invertible family of QFTs with a family anomaly of the form \eqref{noninvertiblefamanomaly} must flow in the IR to 1.) a gapless theory, 2.) a theory which spontaneously breaks $\CG$, or 3.) a non-invertible family of theories with a phase transition.

\section{Examples via deformations of $4d$ $\CN=2$ SYM}
\label{sec:nequals2}

We now illustrate some of the features of categorical family structures in $4d$ $SU(2)$ $\CN=2$ SYM. This $SU(2)$ gauge theory has two adjoint-valued Weyl fermions, in the fundamental of the $SU(2)_R$ global symmetry.  There is a classical, chiral $U(1)_r$ under which the fermions both have $r$-charge 1 and the adjoint-valued complex scalar $\phi$ of the $\CN =2$ gauge multiplet has $r$-charge 2, which is broken $U(1)_r\to \IZ_{8}$ by an ABJ anomaly.   The ABJ anomaly implies that there is no $\theta$ angle in the UV theory: shifts of $\theta$ can be absorbed by fermion field redefinitions.  The $\IZ_8$ chiral symmetry has a mixed anomaly with the $\IZ_2^{(1)}$ center symmetry:
\eq{
\CA\supset \frac{2\pi  }{4}\int A_1\cup \CP(B_2)~,
}
where $A_1$ is the background gauge field for the $\IZ_8$ global symmetry. 
 
The exact, IR low-energy effective theory is obtained in \cite{Seiberg:1994rs,Seiberg:1994aj}.  There is a Coulomb branch moduli space of $\CN =2$ supersymmetric vacua, with the gauge group is Higgsed, $SU(2)\to U(1)$.  This moduli space can be parameterized by the expectation values of $u=\frac{1}{2} \Tr [\phi ^2]$.  The $\IZ _8\subset U(1)_R$ maps $u\to -u$, so $u\neq 0$ spontaneously breaks  $\IZ_{8}\to \IZ_4$.  The IR $U(1)$ gauge theory on the Coulomb branch has a complexified gauge coupling $\tau (u)={da_D\over da}$ of the form
\eq{\label{SU2tau}
\tau(u)\equiv \tau _{IR}=2\tau_{UV}-\frac{4}{2\pi i}\log(u/\Lambda^2)+\sum_{n>0}c_n\frac{\Lambda^{4n}}{u^{2n}}~,
}
where $\Lambda$ is the dynamical scale, the $\log$ term is the one-loop exact perturbative contribution, and the power series are the non-perturbative instanton corrections~\cite{Seiberg:1994rs,Seiberg:1994aj}. The $\IZ_8\subset U(1)_r$ symmetry generator takes $u\to -u$ and $\tau_{UV}=\frac{4\pi i}{g_{UV}^2}+\frac{\theta_{UV}}{2\pi}\to \tau _{UV}+1$.  

Consider the theory far from the origin of the Coulomb branch, $|u|\equiv u_0 \gg |\Lambda |^2$. There one can neglect the instanton contributions and  the IR $\CN =2$, $U(1)$ theory is described by 
\eq{\label{tauIR}
\tau_{IR}\approx 2\tau_{UV}-\frac{4}{2\pi i}\log(u_0/\Lambda^2)+\frac{2\alpha}{\pi}\quad, \quad u=e^{- i \alpha}u_0\in  \IC~.
}
In this regime, the IR theory has an effective, emergent  $\theta_{IR}$ given by 
\eq{
\theta_{IR}=2\theta _{UV} +4\alpha~.
}
As in section~\ref{sec:4dMaxwell}, on a spin manifold the IR theory far from the origin is invariant under
\eq{
Z_{IR}[\theta_{IR}+2\pi]=Z_{IR}[\theta_{IR}]~, 
}
The transformation $\theta_{IR}\mapsto \theta_{IR}+2\pi$ is equivalent to $u\mapsto -iu$ (which is broken by instantons in the full theory).

The theory can also be put on a non-spin manifold, with a generalized $Spin_{SU(2)_R}(4)$ structure \cite{Witten:1994cg,Witten:1994ev,Moore:1997pc,Cordova:2018acb,Wang:2018qoy,Brennan:2023vsa,Brennan:2023kpo,Moore:2024vsd}.  This requires  turning on a $SO(3)_R\subset SU(2)_R$ background gauge field with obstruction class $w_2(R)\in H^2(BSO(3)_R;\IZ_2)$ that is set equal to the second Stiefel-Whitney class of the spacetime $M$:
\eq{
w_2(R)=w_2(TM)\in H^2(M;\IZ_2)~. 
}
The partition function is then not invariant under the shift $\theta_{IR}\mapsto \theta_{IR}+2\pi$ since:
\eq{
\int [d\Psi]~e^{- S_\theta[\Psi]}=\int [d\Psi]~e^{-S_{\theta+2\pi}[\Psi] -\pi i \int \frac{f}{2\pi}\wedge \frac{f}{2\pi}}=\int [d\Psi]~e^{-S_{\theta+2\pi}[\Psi] -\pi i \int \frac{f}{2\pi}\wedge \widehat{w}_2(TM)}~,
}
where $\widehat{w}_2(TM)$ is the integral lift of the second Stiefel-Whitney class and $f$ is the field strength of the dynamical, IR Coulomb branch's $U(1)$ gauge field. We can now use the 1-form magnetic symmetry, whose background gauge field $B_2^{(m)}$ couples as
\eq{
S\supset \itwopi \int B_2^{(m)}\wedge f_2~,
}
 to absorb this additional phase:
\eq{
Z_{IR}[B_2^{(m)};\theta_{IR}]=Z_{IR}[B_2^{(m)}+\pi \widehat{w}_2(TM);\theta_{IR}+2\pi]~.
}
So, as with $U(1)$ electromagnetism, the IR limit of $\CN=2$ $SU(2)$ Yang-Mills theory far out on the Coulomb branch is a higher family of theories characterized by 
\eq{
\theta_{IR}\sim \theta_{IR}+2\pi \quad, \quad B_2^{(m)}\mapsto B_2^{(m)}+\pi \widehat{w}_2(TM)~,
}
where the shift in $\theta_{IR}$ transforms the minimal monopole into a dyon with electric charge 1 (in the convention where the $W$-boson has charge 2), and then the $B_2$ shift cancels the angular momentum in the electromagnetic field carried by the dyon. 

Far from the origin on the Coulomb branch, the IR $U(1)$ gauge theory admits an approximate non-invertible family structure as described in Section~\ref{sec:4dMaxwell}: shifting $\theta\mapsto \theta+2\pi q$, where $q\in \IQ/\IZ$, combined with a discrete gauging.  The fact that the IR $U(1)^{(1)}_m$ 1-form magnetic symmetry is not a symmetry of the UV theory implies that the higher family and non-invertible family structures are broken in the UV. 
 Indeed, the shift $\theta_{IR}\mapsto \theta_{IR}+2\pi q$ corresponds to a phase rotation of only $u$ by $u\mapsto e^{-\frac{\pi i q}{2} u }\phi$ (without shifting $\tau_{UV}$). This is not a symmetry of the UV theory 
(even if embedded into $U(1)_r$)  
 and is indeed explicitly broken in the IR theory by the instanton contributions to the full effective action (described in \eqref{SU2tau}). These contributions break both the higher family and non-invertible family structures. 
 
There are two scales for the emergent family structures. The  1-form magnetic symmetry is broken at the scale of the mass of the monopoles $E_{\rm 1\text{-}form}\sim m_{mono}$. In the semiclassical limit where $u_0\gg \Lambda^2$ and $g_{IR}(u_0)\ll 1$,
\eq{
m_{mono}^2\sim \frac{u_0}{g_{IR}^2(u_0)}~.
}
Following our previous discussion, the family structure is broken at the scale where the instanton corrections become relevant $E_{\rm family}\sim \Lambda$. Using the fact that $\Lambda^2\ll u$,  
we see that $E_{\text{1-form}}\gg E_{\rm family}$ 
and the higher family inequality is obeyed in the semiclassical limit. 

Geometrically, this categorical family structure and its breaking by UV physics follows from the fact that for $u>>\Lambda^2$, the Coulomb branch effectively looks like the classical moduli space which has a single singularity at $u=0$. The classical moduli space is rotationally invariant ($U(1)_r$-invariant) which is equivalent to a shift of $\theta_{IR}$. As we go to strong coupling, we resolve the singularity into the monopole and dyon points at $u=\pm \Lambda^2$ which explicitly breaks $U(1)_r\to \IZ_8$ and therefore breaks both the higher family and categorical family structure.  

\vspace{0.25cm} 

As in~\cite{Seiberg:1994rs,Seiberg:1994aj}, consider breaking $\CN =2$ to $\CN =1$ by adding a superpotential
\eq{
W_{UV}=\frac{m}{2}\, \Tr[\Phi^2]~ \to W_{IR}=mu;}
 the theory then RG flows to $SU(2)$ $\CN=1$ SYM. The deformation explicitly breaks  the $SU(2)_R$ symmetry which allowed us to put the theory on a non-spin manifold and it also explicitly breaks the $U(1)_r \supset \IZ_8$ global symmetry to $U(1)_r\to \IZ_4$. This theory has a parameter $m\, e^{i \theta_{UV}/2}= m\, e^{i \theta_{IR}/4}$ whose phase is a spurion for the broken $U(1)_r$ symmetry. 
 
 The IR theory has two $\CN =1$ supersymmetric vacua at $u=\pm \Lambda^2$~\cite{Seiberg:1994rs}. 
These vacua are related by $u\mapsto e^{\pi i }u$, which corresponds to $\theta_{IR}\mapsto \theta_{IR}+4\pi$, and is the explicitly broken generator of $\IZ_8$. These $\CN=1^\ast$ theories thus fit into an ordinary family of theories parametrized by $m\,e^{i \theta_{UV}/2}$ with $m\sim -m$. Since the $\IZ_8$ has a mixed anomaly with the $\IZ_2^{(1)}$ center symmetry of the UV theory, the two vacua are equivalent and the domain wall interpolating between them has a world volume anomaly given by $\omega_d=\frac{\pi }{2}\CP(B)$. In the $\CN=1$ deformed theory, there is no IR higher family or categorical family structure. 

\subsection{$SU(2)$ vs. $SO(3)$ $\CN=2$ SYM}
\label{sec:SU2SO3}
Gauging the $\IZ_2^{(1)}$ center symmetry of the $SU(2)$ $\CN=2$ SYM theory yields the $\CN=2$ $SO(3)\cong SU(2)/\IZ _2$ Yang-Mills theory.  
   The $SO(3)$ theory has a similar vacuum structure and low-energy effective theory\footnote{As discussed in~\cite{Seiberg:1994aj} in $SU(2)$ it is natural to take the adjoint $W$ bosons to have $U(1)$ electric charge 2, whereas in~\cite{Seiberg:1994rs} they were taken to have $U(1)$ electric charge 1 -- which is natural for $SO(3)$.  Although gauging $\IZ_2^{(1)}$ does not affect the local physics, this difference in the normalization results in the ``isogeny" for the $SO(3)$~\cite{Seiberg:1994rs} vs $SU(2)$ with $N_f=0$ curves, as discussed in~\cite{Seiberg:1994aj}. See also~\cite{Closset:2023pmc} for recent extensions.}, with a Coulomb branch parametrized by $\langle u\rangle\in \IC$ with $u=\half Tr\,[\phi^2]$, though now the massless monopole / dyon singularities, at $u=\pm \Lambda^2$, are physically inequivalent.  Since $SO(3)$ gauge theory has $1/4$-instantons on generic manifolds, the UV theory now has a non-invertible $\IZ_8\subset U(1)_r$ global symmetry that contains both a higher-group global symmetry for the $\IZ_4\subset \IZ_8$ and a 0-form global symmetry for the $\IZ_2\subset \IZ_8$.  Since the $\IZ_8$ symmetry of the $SU(2)$ theory generated the IR family relation, our discussion in Section \ref{sec:gauging} implies that the $SO(3)$ $\CN=2$ SYM will have an emergent categorical family structure that is preserved by the instanton corrections.

One can obtain the IR theory of $SO(3)$ $\CN=2$ SYM by starting with the IR theory of $SU(2)$ $\CN=2$ SYM and gauging the discrete $\IZ_2^{(1)}$ center symmetry.  It is convenient to rescale the $U(1)$ gauge field so that the minimal Wilson and monopole lines are the integer quantized lines of the new $U(1)$ gauge field:
\eq{
\frac{F_{U(1)}}{2\pi}\longmapsto \half \frac{\tildeF_{U(1)}}{2\pi}=\half w_2(\IZ_2)~{\rm mod}_\IZ~,
}
where $w_2(\IZ_2)\in H^2(M;\IZ_2)$ is the obstruction class of lifting the $SO(3)$ gauge bundles to $SU(2)$ gauge bundles. The $\theta$-angle for the rescaled gauge group $\tilde\theta_{IR}=\frac{\theta_{IR}}{4}$ now shifts by $\tilde\theta_{IR}\mapsto \tilde\theta_{IR}+\half\alpha$ under the shift $\theta_{UV}\mapsto \theta_{UV}+\alpha$.  This will lead to two generalized family structures which are matched by the UV non-invertible $\IZ_8$ structure:
\eq{
\text{Higher Family Structure :}&~\tilde\theta_{IR}\longmapsto \tilde\theta_{IR}+2\pi~,~B_2^{(m)}\mapsto B_2^{(m)}+\pi w_2(TM)~,\\
\text{Non-Invertible Family Structure :}&~\tilde\theta_{IR}\longmapsto \tilde\theta_{IR}+\pi ~,~\CS\text{ operation}~.
}

If we deform by the $\CN=1$ preserving mass term, there will be a family of theories parametrized by $m\,e^{i \theta_{UV}/2}$. This deformation  explicitly breaks the  non-invertible symmetry $\IZ_8\to \IZ_4$ and also breaks the $SU(2)_R$ global symmetry; this restricts consideration to spin manifolds, where the higher family structure is trivial. 
The $\CN=1$ deformation of the $SO(3)$ theory again lifts the Coulomb branch except for the two gapped vacua where the massless monopoles or dyons condense.  These two theories, differing by $\Delta \theta _{UV}=2\pi$ and thus $\Delta \tilde\theta _{IR}=\pi$, are physically inequivalent:  one is trivially gapped and the other has a  $\IZ_2$ gauge theory~\cite{Aharony:2013hda}.  The two inequivalent vacua are related by a non-invertible family relation. 

For the vacuum with a  $\IZ_2$ gauge theory, say the monopole vacuum of the $SO(3)_+$ theory, the $\IZ _2$ gauge theory TQFT can be described by the action 
\eq{
S_{IR}\supset \frac{2i}{2\pi}\int da_1\wedge b_2~. 
}
In terms of the physics of the UV theory, the $a_1$ is the ``magnetic'' gauge field and   hence $\frac{2 b_2}{2\pi}=w_2(\IZ_2)~{\rm mod}_2$ is identified with the discrete flux $w_2(\IZ_2)$ that is summed over in the path integral.  Thus shifting $\theta _{UV}\to \theta _{UV}+2\pi$, which exchanges $SO(3)_\pm \leftrightarrow SO(3)_\mp$ and exchanges the two vacua of the IR $\CN =1$ SYM $SO(3)$ theory~\cite{Aharony:2013hda}, shifts the above action by 
\eq{
\Delta S=2\pi i\times \frac{1}{4} \int \frac{2b_2}{2\pi}\wedge \frac{2b_2}{2\pi}=\frac{i}{2\pi}\int b_2\wedge b_2~.
}
This trivializes the TQFT. 

In the low energy  $\CN=1^\ast$ $SO(3)$ theory, the non-invertible family relation is generated by rotating $m\mapsto -m$ supplemented by a discrete gauging $\CS$. The phase rotation of $m$ can be absorbed by a field redefinition by the anomalous $\IZ_8$ rotation. This effectively shifts $\theta_{UV}\mapsto \theta_{UV}+2\pi$. To demonstrate the non-invertible family structure, we now show that the $\CS$ gauging cancels this shift.  The $\CS$ gauging is implemented by the action 
\eq{
S\longmapsto \int\frac{2i}{2\pi} (da_1\wedge b_2+dc_1\wedge B_2)+\itwopi B_2\wedge B_2+\frac{2 i}{2\pi}b_2\wedge B_2~.
}
Integrating out $c_1$ projects to a sum over $B_2\in H^2(X_d;\IZ_2)$ and we obtain
\eq{
S\longmapsto \itwopi \int 2da_1\wedge b_2+(B_2+b_2)\wedge (B_2+b_2)-b_2\wedge b_2~.
}
Integrating out $B_2$ then gives 
\eq{
S\longmapsto \frac{2i}{2\pi}\int da_1\wedge b_2-\itwopi \int b_2\wedge b_2~.
}
This agrees with the result of the previous paragraph.  There is thus indeed a non-invertible family structure, with the  gapped theories at the two different vacua given by  
\eq{
S_{\IZ_2}=\frac{2i}{2\pi}\int da_1\wedge b_2\quad, \quad \widetilde{S}_{\IZ_2}=\frac{2i}{2\pi}\int da_1\wedge b_2+\itwopi\int b_2\wedge b_2~. 
}

Consider first the theory $\tildeS_{\IZ_2}$. As discussed in \cite{Kapustin:2014gua}, for the theory to be well-defined the gauge transformations must be modified as 
\eq{
\delta b_2=d\Lambda_1\quad, \quad \delta a_1=d\lambda-\Lambda_1~. 
}
The Wilson line operators of $a_1$ are then not gauge invariant and must bound $b_2$ sheets:
\eq{
W(\gamma)=e^{i \oint_\gamma a_1}~\longmapsto ~\tildeW(\gamma;\Gamma)=e^{i \oint_\gamma a_1+i \int_\Gamma b_2}\quad, \quad \partial \Gamma=\gamma~. 
}
Conversely, any $b_2$ surface operator $\CS(\Sigma)=e^{i \oint_\Sigma b_2}$ can be trivialized by cutting a hole and transforming it into a $\tildeW$ operator and then contracting it to a point. Since there are no non-trivial operators the theory $\tildeS_{\IZ_2}$ is the trivial theory, matching with the expectation that the dyon point of $SO(3)_+$ $\CN=1$ theory is a trivially gapped IR theory \cite{Aharony:2013hda}. 

Note that shifting $\theta_{UV}\mapsto \theta_{UV}+4\pi$ is a trivial transformation: shifting by the above family relation twice gives the shift 
\eq{
\CS^2:S\longmapsto \frac{2i}{2\pi}\int da_1\wedge b_2-\pi i \int \CP\left(\frac{2b_2}{2\pi}\right)~. 
}
On shell, $b_2$ is a $\IZ_2$-quantized field ($e^{i\oint b_2}=\pm1$). Since the $\CN=1^\ast$ theory restricts us to spin manifolds, the second term evaluates $e^{\pi i \int \CP\left(\frac{2b_2}{2\pi}\right)}=1$ and is therefore trivial. 

As in~\cite{Aharony:2013hda}, the above discussion of the line and surface operators can be interpreted in terms of  confinement or oblique confinement via monopole or dyon condensation (see e.g.~\cite{tHooft:1981bkw,Cardy:1981qy,Cardy:1981fd,Seiberg:1994rs,Intriligator:1995id} for additional discussion).   For one of the vacua, 
the IR physics is governed by charge 2 monopole condensation, Higgsing the $U(1)$ gauge theory of the $\CN =2$ Coulomb branch to $\IZ _2$ gauge theory, with charge 1 monopole operators unscreened.  The physics of the other vacuum, where the dyons are massless and condense, can be regarded from the IR perspective as an effect of the  $\Delta \theta _{IR}=\pi$ shift, since the monopole lines then acquire electric charge via the Witten effect and then must have attached $b$-surfaces, which confine the line operators in the topological theory. The $SO(3)_+$ vs $SO(3)_-$ cases differ only by an exchange of the roles of the two vacua.  The distinct confining vs oblique confining IR phases of the two inequivalent vacua are  related by the non-invertible family relation.

\section{Families of Deformations of non-Abelian Gauge Theories}
\label{sec:dangerouslyirrelevant}

We now consider categorical families of theories in the context of some irrelevant deformations of 4d non-abelian gauge theories.  We consider IR-irrelevant deformations to keep the IR theory under control (though in some cases the deformations we consider will turn out to be dangerously irrelevant).   The theory with the irrelevant deformation can be regarded as the low-energy effective field theory of a  UV completion along some RG flow.

\subsection{$4d$ $SU(N_c)$  $\CN=1$ SYM with SUSY-Breaking, Irrelevant Deformations}

Consider $4d$ $SU(N_c)$ Yang-Mills with a single adjoint Weyl fermion $\lambda$. For $\lambda$ massless, the IR theory is $\CN=1$ super-Yang-Mills, and for massive $\lambda$ the IR theory is $\CN =0$ Yang-Mills; we here consider the massless case.  
This theory has a global $\IZ_{N_c}^{(1)}$ center symmetry and a $\IZ_{2N_c}^{(0)}$ chiral symmetry with mixed  anomaly~\cite{Gaiotto:2014kfa, Hsin:2018vcg}
\eq{
\CA=\frac{2\pi  }{{\rm gcd}(N_c, 2)N_c}\int a_1\cup \CP(B_2)~.  
}
This anomaly prevents the theory from being symmetry-preserving and trivially gapped in the IR~\cite{Cordova:2019bsd}. This fits with the supersymmetric analysis in~\cite{Witten:1982df,Affleck:1983mk,Amati:1988ft,Seiberg:1994bz,Seiberg:1994rs} and references therein, which show that the theory confines in the IR with a mass gap, unbroken $\IZ_{N_c}^{(1)}$,  and $\IZ_{2N_c}^{(0)}\to \IZ _2^{(0)}$ spontaneous symmetry breaking by gaugino condensation $\langle \lambda \lambda \rangle=|\Lambda|^3 \times e^{2\pi i k/N_c}$.  This all fits with the counting of vacua by the $T^3$ Witten index, ${\rm Tr}(-1)^F=N_c$~\cite{Witten:1982df}. 

We now deform the theory by the supersymmetry-breaking interaction
\eq{\label{1AQCDPotential}
\Delta \CL =m_n (\lambda \lambda)^n+c.c. 
}
taking $n>1$ to exclude a mass term.  Non-zero $m_n$ explicitly breaks $\IZ_{2N_c}\to \IZ_{2r}$ where $r={\rm gcd}(N_c,n)$. The continuous family of theories parametrized by the phase $\theta$ of the coupling $m_n=e^{i \theta}|m_n|$  has an anomaly. To see it, note that $\theta$ can instead arise for real $m_n=|m_n|$ via a phase rotation of $\lambda$, $\lambda\mapsto e^{\frac{i \theta}{2n}}\lambda$, which by the ABJ anomaly shifts $\Delta \theta _{YM}=N_c\theta /n$:
\eq{
\Delta S=\frac{2N_ci \,\theta}{2n}\int \frac{\Tr[F\wedge F]}{8\pi^2}~. 
}
The phase $\theta$ of $m_n$ thus exhibits the periodicity  
\eq{
\theta \sim \theta+\frac{2\pi n}{N_c}=\theta+\frac{2\pi \ell}{L}~,
}
where $N_c=rL$ and $n=r\ell$ and for simplicity we assume that ${\rm gcd}(L,2r)=1$.   
A family anomaly in the $\theta$ periodicity arises upon turning on a background gauge field for the $\IZ_{N_c}^{(1)}$ global symmetry,  as
(where we reduce by the anomalous $\IZ_{2r}$ global symmetry) 
\eq{
Z_{\theta+\frac{2\pi \ell}{L}k}[B_2]=Z_{\theta}[B_2]\times e^{2\pi i  k\frac{N_c-1}{L}\int \CP(B_2)}~,
} 
This phase can be written in terms of the family anomaly
\eq{
\CA=\frac{N_c-1}{\ell}\int d\theta\cup \CP(B_2)=\frac{N_c-1}{L}\int d\Theta\cup \CP(B_2)~, 
}
where in the last expression we rescaled $\Theta=\frac{N_c}{n}\theta=\frac{L}{\ell}\theta$ so that $\Theta\sim \Theta+2\pi$.   

As in~\cite{Cordova:2019uob}, the above family anomaly rules out the IR scenario of a symmetry preserving, gapped IR phase without phase transitions for the $m_n$ deformed theory\footnote{The theory also has the analog of the mixed anomaly of~\cite{Gaiotto:2017yup} involving time reversal symmetry, here at $\theta=\frac{\pi k\ell}{L}$ where $k\in \IZ$. The phase transitions must allow matching of that anomaly too.}.  
Since the interaction we are considering is irrelevant, it {is reasonable to assume} that the theory still confines and flows to a gapped phase; 
we do not expect that the anomaly is matched with a gapless IR phase, nor with an IR phase with the 1-form center symmetry spontaneously broken (i.e. non-confining).  As in~\cite{Cordova:2019uob}, we expect that the anomaly is matched by a phase transition in between $\theta,\theta+\frac{2\pi \ell}{L}$.   The anomaly implies that there are $L$ phase transitions at special values of $\theta$.

The irrelevant deformation~\eqref{1AQCDPotential} indeed explicitly breaks $\IZ_{2N_c}\to \IZ_{2r}$ and consequently generically lifts some of the $N_c$-vacua that arose from spontaneously breaking $\IZ_{2N_c}\to \IZ_2$.   There are then $r$ vacua, which are protected by the spontaneous breaking of $\IZ_{2r}\to \IZ_2$. The original vacua are broken up into $L$ sets of $r$ vacua, and for generic values of $\theta$, only one such set of $r$ are the vacua of the theory. The $L$ sets of $r$ vacua are then exchanged as the phase of $m$ is rotated through $2\pi$ There is level crossing between the sets of $r$ vacua at each $\theta=\frac{\pi (2k+1)\ell}{L}$ with $k\in \IZ$, matching also the time-reversal anomaly there. 

This scenario also follows if we treat the interaction \eqref{1AQCDPotential} as a perturbation of 
the effective field theory for the phase of the fermion condensate (taking $m_n=|m_n|$ real by phase rotation of $\lambda$). In this approximation, the irrelevant interaction leads to a potential 
\eq{
V\sim m_n \langle \lambda\lambda\rangle^n+c.c.\longmapsto V_k(\theta)=|m_n|\,|\Lambda|^{3n}\cos\left(\frac{2\pi k\ell}{L}+\theta\right)\quad, \quad k=0,...,N_c-1~.
}
Since $V_k(\theta)=V_{k+L}(\theta)$, for generic $\theta$ there are $r$ minima of $V_{k}(\theta)$ for $k\in \IZ_{N_c}$, with fixed $k_L=k~{\rm mod}_L$. Since $V_k\left(\theta+\frac{2\pi \ell}{L}\right)=V_{k+1}(\theta)$,  
the $r$-vacua with fixed $k_L$ are 
permuted by shifting $\theta\mapsto \theta+\frac{2\pi \ell}{L}$: $k_L\mapsto k_L-1$.

\subsubsection{Deformations of $SU(N_c)/\IZ_L$ $\CN=1$ SYM}
\label{sec:noninvertibleN1}

Gauging a $\IZ _L^{(1)}\subset \IZ _N^{(1)}$ subgroup of the center symmetry for the theory of the previous subsection yields
$4d$, $\CN =1$ susy $SU(N_c)/\IZ_L$ Yang-Mills. Writing $N_c=rL$, the $\IZ^{(0)}_{2N_c}$ chiral symmetry of the previous subsection is transmuted into a $\IZ_{2N_c}$ non-invertible symmetry that includes a $\IZ_{2r}^{(0)}$ group symmetry. The non-invertible symmetry is  spontaneously broken to $\IZ_2$ by the $\langle \lambda \lambda \rangle$ condensate and leads to $N_c$ vacua.  As in the $SO(3)=SU(2)/\IZ _2$ case, some of the vacua are related by non-invertible symmetries and non-trivially gapped  
since the non-invertible symmetry defect induces a discrete gauging TQFT. 

Now consider deforming the $\CN =1$ theory by the susy-breaking irrelevant operator
\eq{
\Delta \CL=m_n (\lambda \lambda)^n+c.c. ~. 
}
If ${\rm gcd}(N_c,n)=L$, this deformation completely breaks only the group-like global symmetry and the result follows straightforwardly from the discussion in the previous subsection: the IR theory has generically $L$ vacua which are related by non-invertible symmetry defects and there are $r$ phase transitions that follow from the standard family anomaly. Again, each of these vacua is described by a $\IZ_L$ gauge theory with a different discrete theta term.   On the other hand, if the deforming operator above has ${\rm gcd}(N_c,n)=r$, the deformation explicitly breaks the non-invertible symmetry.  This leads to $r$ vacua for generic values of $m_n$, related by the spontaneously broken $\IZ_{2r}$ group-like symmetry operators.  In order to probe the vacuum structure of this theory, we need to realize that the phase of $m_n$ participates in a categorical family structure.  Indeed, the phase rotation shifts the integrand of the partition function as
\eq{
m_n\longmapsto e^{i \theta} m_n\quad\Longrightarrow \quad \Delta S=\frac{i \theta }{2n}(2N_c)\int \frac{\Tr[F\wedge F]}{8\pi^2}=i \theta \frac{L}{\ell}\int \frac{\Tr[F\wedge F]}{8\pi^2}~,
}
where $n=\ell r$. 
Since
\eq{
\int \frac{\Tr[F\wedge F]}{8\pi^2}=\frac{r(N_c-1)}{2L}\int \CP(w_2)~\qquad\rm{mod} \ \IZ~, 
}
where $w_2\in H^2(BSU(N_c)/\IZ_L;\IZ_L)$ is summed over in the path integral, $\theta$ is periodic $\theta\sim \theta+2\pi \ell$; this is generated by the $\IZ_{2r}$ group action on $\lambda$. 

Note that $\theta\sim \theta+\frac{2\pi \ell}{L}$ will also lead to a categorical family structure, with a family anomaly, since shifting $\theta\mapsto \theta+\frac{2\pi p\ell}{L}$ shifts the action by
\eq{
\Delta S&
=2\pi i p \frac{L(N_c-1)}{2r}\int \CP(B_2)~,
}
where we have assumed $r,L$ are co-prime.  Again, this anomaly can be matched via phase transitions in the family of theories between $\theta$ and $\theta+\frac{2\pi p\ell}{L}$.  This fits with the picture of  the $N_c$ vacua deforming into sets of $r$ degenerate vacua, with $L$ phase transitions as $\theta$ is rotated. 

\subsection{$4d$ $SU(N_c)$ with $N_f=2$ Adjoint QCD}

Consider $4d$, non-supersymmetric, $SU(N_c)$ Yang-Mills with $N_f=2$ adjoint-valued Weyl fermions $\lambda ^f_\alpha$, with $\alpha =1,2$. This theory has global symmetry given by 
\eq{
G=\IZ_{N_c}^{(1)}\times \frac{SU(2)_R\times Spin(4)\times \IZ_{4N_c}}{\IZ_2\times \IZ_2'}~,
}
where $\IZ_{4N_c} \subset U(1)_r^{(0)}$ is a chiral symmetry, and the two $\IZ_2$ quotients are generated by 
\eq{
\IZ_2:~(-1)^F\circ (-\mathds{1}_{SU(2)_R})\quad, \quad \IZ_2':~(-\mathds{1}_{SU(2)_R})\circ (-1_{\IZ_{4N_c}})~.
} 
This theory has a mixed $\IZ_{N_c}^{(1)}\times \IZ_{4N_c}$ anomaly and mixed $\IZ_{4N_c}\times SU(2)_R^2$ anomaly:
\eq{
\CA\supset 2\pi \frac{N_c-1}{2N_c}\int a_{4N_c}\cup \CP(B_2)+2\pi  \frac{(N_c^2-1)}{4N_c}\int a_{4N_c}\wedge\frac{ \Tr[F_R\wedge F_R]}{8\pi^2}~,
}
where $B_2,~a_{4N_c}$, and $F_R$ are the background gauge fields for $\IZ_{N_c}^{(1)},\IZ_{4N_c}^{(0)}$ and the field strength of the background gauge field for $SU(2)_R$ respectively and we are using the convention $\oint a_{4N_c}=0,1,...,4N_c-1$. 
For $N_c=4n_c+2$, this theory also has a discrete anomaly~\cite{Cordova:2018acb,Brennan:2023vsa}:
\eq{
\CA\supset \pi  \int B_2 \cup w_3(TM)~.
}

A plausible scenario for the IR dynamics is confinement and $SU(2)_R\to U(1)$ spontaneous symmetry breaking from fermion condensation of $ {\cal O}^{(fg)}=\Tr[\lambda^{(f}\lambda^{g)}]$, leading to $N_c$ copies of the $\ICP^1$ sigma model. See in particular~\cite{Cordova:2018acb,DHoker:2024vii} for discussion and how this IR phase arises via a deformation of $\CN =2$ SYM.   The $\ICP^1$ fields are the Goldstone fields from spontaneously breaking $SU(2)_R\to U(1)$ and the $N_c$ copies come from spontaneously breaking $\IZ_{4N_c}\to \IZ_2$\footnote{ 
There are only $N_c$ copies; the extra broken $\IZ_2\subset \IZ_{4N_c}/\IZ_2$ acts on the  $\ICP^1$ Goldstone fields~\cite{Cordova:2018acb,DHoker:2024vii}.}. 

We now consider the effect of deforming the $N_f=2$ adjoint fermion theory by higher dimension fermion composite operators. We first consider operators that preserve the $SU(2)_R$ global symmetry, but may break the chiral symmetry, e.g.  quartic operators like the double-trace ${\cal O}^{(fg)}{\cal O}^{(fg)}$. For $N_c\geq 4$, we can also consider single-trace, $SU(2)_R$ preserving 4-fermion operators using the quartic Casimir; we will not here distinguish these two operators and will simply write both as $\lambda \lambda \lambda \lambda$.  We can consider deformations by such operators and higher dimension operator versions:
\eq{
\Delta \CL =m_n(\lambda\lambda\lambda\lambda)^n+c.c.~. 
}
This interaction breaks $\IZ_{4N_c}\to \IZ_{4r}$ where $N_c=rL$, $n=r\ell$ and the phase of $m_n=e^{i \theta}|m_n|$, which has periodicity 
 $\theta\sim \theta+\frac{2\pi \ell}{L}$, 
has an anomaly with the $\IZ_{N_c}^{(1)}$ center symmetry:
\eq{
\CA\supset \frac{N_c-1}{2n}\int d\theta \cup \CP(B_2)~.
}

This anomaly fits with having the $N_c$ $\ICP^1$ NLSMs of the $N_f=2$ theory perturbed by the $m_n$ deformation into $L$ sets of $r$ $\ICP^1$ NLSMs where for generic values of $\theta$ there are $r$ degenerate copies in the groundstate, with the others lifted above in energy.   The above family anomaly again implies that there is a phase transition as $\theta\mapsto\theta+ \frac{2\pi \ell}{L}$ where there is a ``level crossing'' of NLSMs. The effect of the $\Delta \CL$ deformations is to introduce a small gap between the $N_c=rL$ copies of the NLSM, which are exchanged as one dials the phase of $m_n=e^{i \theta}|m_n|$.  If one of the fermions is given a mass, the degeneracy of each of the $\CP ^1$'s is lifted and this RG flows to fit with the above discussion of the $N_f=1$ adjoint fermion case. 

\subsubsection{$SU(2)_R$-Breaking Deformation}

We now instead consider deforming the $N_f=2$-adjoint fermion theory by deformations
\eq{\label{2adjointSU2Rbreaking}
\Delta \CL =m_n (\lambda\lambda)^{2n+1}+c.c.~.
}
This deformation is  irrelevant for $n>0$ and breaks both the chiral symmetry $\IZ_{4N_c}\to \IZ_{2r}$ where ${\rm gcd}(N_c,2n+1)=r$ and the $(\lambda \lambda )^{2n+1}$ operator is in the adjoint of $SU(2)_R$ so $m_n$ explicitly breaks $SU(2)_R\to U(1)_R$.   The explicit breaking of $SU(2)_R$ generates a potential in each $\ICP ^1$ NLSM that localizes to the $U(1)_R$ fixed points -- the north and south poles -- so the deformation is dangerously irrelevant. Since the broken generator of $\IZ_4\subset \IZ_{4N_c}$ acts as reflection in the $\ICP^1$ NLSM, the deformation also lifts the degeneracy between the north and south poles, leading  to a gapped theory with vacua parametrized by the spontaneous breaking of $\IZ_{2r}\to\IZ_2$. As we will now discuss, the family anomaly in  the phase of the coupling constant $m_n$ suggests additional structure to the IR vacua phases.

For $N_c=4n_c+2$, this theory also has a $\IZ _2$ non-perturbative anomaly 
\eq{
\CA\supset \pi  \int B_2\cup w_3(TM)~, 
}
where $B_2$ is a background gauge field for the $\IZ_2^{(1)}\subset \IZ_{4n_c+2}^{(1)}$ center symmetry and $w_3(TM)$ is the Stiefel-Whitney class which can be activated by coupling the theory to a $Spin_{SU(2)_R}(4)$ global symmetry structure (as discussed in \cite{Brennan:2023vsa,Cordova:2018acb,Wang:2018qoy,Moore:2024vsd}). 
This anomaly is computed by turning on the background $B_2$ gauge field and putting the theory on $\ICP^2$ with a $Spin_{SU(2)_R}(4)$ structure and then looking at the action of a particular bosonic symmetry 
\eq{
\widehat\varphi=\varphi_{c.c.}\circ W_{SU(4n_c+2)}\circ W_{SU(2)_R}~,
}
where $\varphi_{c.c.}$ is the complex conjugation action on $\ICP^2$ and $W_G$ is a particular $\IZ_2$ Weyl element which obeys $W_G^2=(-\mathds{1}_G)\subset Z(G)$.  The partition function transforms as
\eq{
\widehat\varphi:Z\longmapsto Z\times e^{\pi i \int B_2\cup w_2(TM)}~. 
}
The fermion bi-linear transforms as
$\widehat\varphi(\lambda\lambda)=-\lambda\lambda$.

In the case where we add the $SU(2)_R$-breaking interaction as in \eqref{2adjointSU2Rbreaking}, we can match the $\widehat\varphi$ action by replacing 
the action of $W_{SU(2)_R}$ with charge conjugation \cite{Brennan:2023vsa} which in the IR $\ICP^1$ NLSM is enacted by a $\pi$-rotation which exchanges the north and south poles.  It is clear from the action of $\widehat\varphi$ on on $\lambda\lambda$ that the $\Delta \CL$ interaction breaks the $\widehat\varphi$ symmetry.  Given the action of $\widehat\varphi$ in the IR $\ICP^1$ theory, the north and south poles of a given $\ICP^1$ NLSM will not be degenerate ground states. The action of $\widehat\varphi$ on $\lambda\lambda$ implies an anomalous family identification 
\eq{
m_n\sim -m_n\quad \Longrightarrow \quad \Delta S=\pi i  \int B_2\cup w_2(TM)~,
}
with a family anomaly for the phase of  $m_n=e^{i \theta}|m_n|$:  
\eq{
\CA_{fam}= \int d\theta \cup B_2\cup w_2(TM)~.
} 

Moreover, for $m_n=e^{\frac{\pi i k}{2}} |m_n|$ with $k\in \IZ$, there is an additional time reversal symmetry involving the action of $\widehat\varphi$ with a mixed anomaly
\eq{
\CA\supset \pi  k\int a_{\widehat\varphi}\cup B_2\cup w_2(TM)~,
}
where $a_{\widehat\varphi}$ is the background gauge field for the $\widehat\varphi$ action. 
This  implies that the theory must have a degenerate ground state at $m_n=\pm i |m_n|$. 
The IR theory can be generically gapped, where the vacua corresponds to the expectation value of the $\ICP^1$ Goldstone at one of the two $U(1)_R\subset SU(2)_R$ fixed points.  There will be phase transitions in the IR theory as the phase of $m_n$ winds through $m_n=\pm i |m_n|$, where the north and south pole fixed points become degenerate and are exchanged by the action of $\widehat\varphi$.  
Combining this discussion with that of the explicit breaking of $\IZ_{4N}\to  \IZ_{2r}$ in the previous section suggests that, as the phase of $m_n=e^{i\theta}|m_n|$ varies $\theta=0\mapsto 2\pi$, it first leads to $L$ level crossings of $r$ trivially gapped vacua at one pole, and then $L$ level crossings of $r$ trivially gapped vacua at the other pole. 

\subsubsection{$SU(4n_c+2)/\IZ_2$ with $SU(2)_R$ Breaking}

\label{sec:2adjgauging}
$SU(4n_c+2)/\IZ_2$ Yang-Mills with 2-adjoint Weyl fermions  has $\IZ_2^{(1)}\times \IZ_{2n_c+1}^{(1)}$ 1-form global symmetry.  The $\IZ_{4N_c}$ chiral symmetry is transmuted into a higher group chiral symmetry that includes a $\IZ_{2N_c}$ group symmetry. This differs from the generic case where we gauge a $\IZ_L^{(1)}\subset \IZ_{N_c}^{(1)}$ because for $L=2$, the anomalous phase activated by the generator of $\IZ_{4N_c}$ is
\eq{
\Delta S=\pi i \int \CP(w_2(\IZ_2))=\pi i \int w_2(\IZ_2)\cup w_2(TM)~,
}
which can be canceled by a Green-Schwarz mechanism.  The $\IZ_{4N_c}$ is then transmuted into a 2-group involving 1-form magnetic symmetry. 
The anomaly of $SU(4n_c+2)$ 2-adjoint QCD:
\eq{
\CA\supset \pi  \int B_2\cup w_3(TM)~,
}
implies that the $\IZ_2^{(1)}$ magnetic 1-form symmetry forms a  2-group with the $Spin_{SU(2)_R}(4)$ global symmetry.

We now consider the theory with the added deformation 
\eq{
\Delta \CL=m_n (\lambda\lambda)^{2n+1}+c.c.
}
Much of the discussion from before is the same except that now the family anomaly associated with $m_n\mapsto -m_n$ under the action of $\widehat\varphi$ is transmuted into a higher family of theories. This can be seen from the fact that the $\widehat\varphi$ action leads to the shift of the integrand of the path integral which is equivalent to shifting the action:
\eq{
\Delta_{\widehat\varphi}S=\pi i \int w_2(\IZ_2)\cup w_2(TM)~,
}
where $w_2(\IZ_2)$ is the $\IZ_2$-valued flux associated to the $\IZ_2$ quotient of the gauge group $SU(4n_c+2)/\IZ_2$. This anomalous phase can be absorbed by the higher family relation
\eq{
m_n\longmapsto -m_n\quad, \quad \widetilde{B_2}\longmapsto \tildeB_2+\pi w_2(TM)~.
} 
This is consistent with the fact that the $\IZ_2$ involution of the $\ICP^1$ model generated by an element of the $\IZ_{4N_c}$ chiral symmetry is transmuted into a higher group involving the 1-form magnetic symmetry upon gauging $\IZ_2^{(1)}\subset \IZ_{4n_c+2}^{(1)}$. 

The $SU(4n_c+2)/\IZ_2$ theory has a vacuum structure that is similar to the $SU(4n_c+2)/\IZ_L$ gauge theory with a single adjoint fermion. There is a family of theories, with phase transitions where the vacua are related by a spontaneously broken non-invertible symmetry, or a non-invertible family of theories with phase transitions where the vacua are related by a spontaneously broken global symmetry as described in Section \ref{sec:noninvertibleN1}.

\subsection{$\CN=2$ $SU(3)$ SYM with $\CN =1$ SUSY-Breaking $\Delta W$ Deformations}

We here consider $\CN=2$ $SU(3)$ Yang-Mills, and  $\CN =2 \to \CN =1$ superpotential deformations 
\eq{\label{eq:DeltaW}
\Delta W=g_2u+g_3 v~, \qquad u=\frac{1}{2}\Tr[\Phi^2], \quad v=\frac{1}{3}\Tr[\Phi^3],
}
where $\Phi$ is the $\CN=1$ chiral superfield component of the $\CN=2$ vector-multiplet, and $g_2,g_3$ are $\CN =2 \to \CN =1$ breaking interactions.  The undeformed $\CN=2$ SYM theory has a $SU(2)_R\times \IZ_{12}$ global R-symmetry group where the  classical $U(1)_r$ is broken to $U(1)_r\to \IZ_{12}$ by an ABJ anomaly. The $\CN=1$ chiral-multiplet $\Phi$ in the $\CN=2$ vector-multiplet has charge 2 under the $\IZ_{12}$ symmetry. This $\IZ_{12}$ has a mixed anomaly with the $\IZ_3^{(1)}$ center symmetry:
\eq{\label{ADZ3anomaly}
\CA=\frac{2\pi }{3}\int a_1\cup \CP(B_2)~.
}

The undeformed $\CN =2$ $SU(3)$ SYM theory has a low-energy theory given by the $\CN =2$ $U(1)^2$ gauge theory with a moduli space of vacua given by expectation values of the operators $u$ and $v$.   Near the singularities of the SW curve, there are light (magnetically) charged 
hypermultiplets $(M_{1,2},\tildeM_{1,2})$ for the IR $U(1) ^2$ gauge fields.  These hypermultiplets couple to the Coulomb branch moduli via the superpotential
\eq{
W_{\CN =2}=A_1(u,v) \tildeM_1 M_1+A_2(u,v)\tildeM_2 M_2~.
}
The functions $A_I(u,v)$ are obtained from the periods of the $\CN =2$ curve.  The structure of the massless monopoles and dyons was analyzed in detail in~\cite{Argyres:1995jj}, as we briefly review here, along with the effect of the $\CN =1$ preserving deformation in \eqref{eq:DeltaW}.    For $g_2\neq 0$ and $g_3=0$, the IR theory is $\CN =1$ $SU(3)$ SYM, which has $\Tr (-1)^F=3$ supersymmetric vacua.  These three vacua are the three places where $A_1=A_2=0$ for two mutually-local monopole / dyon fields.   These three points have $v=0$ and are fixed points of the $\IZ_4\subset\IZ_{12}$, and they are exchanged by the action of $\IZ_3\subset \IZ_{12}$.

For $g_2g_3 \neq 0$, there are five $\CN =1$ supersymmetric vacua: three with mass gap, and two with a massless $\CN =1$ $U(1)$ gauge multiplet~\cite{Argyres:1995jj}. This can also be understood as in~\cite{Elitzur:1996gk} by noting that $\Delta W$ classically yields  two supersymmetric vacua: one at $\Phi =0$, where the IR theory is $\CN =1$ $SU(3)$ Yang-Mills; and one with $\Phi \neq 0$  where the IR theory is $\CN =1$ with $SU(3)\to SU(2)\times U(1)$.  In the quantum theory, the $\CN =1$ $SU(3)$ vacuum near $\Phi =0$ splits into three $\CN =1$ vacua with confinement and a mass gap, where two sets of mutually local monopoles / dyons condense.   These are the only vacua for $g_3=0$.  For $g_{2,3}\neq 0$, there are also the classical $SU(2)\times U(1)$ vacua, which split in the quantum theory into two vacua associated with $\CN =1$ $SU(2)$ YM, where only one massless monopole / dyon condenses. 

Breaking to $\CN =1$ by the deformation  $\Delta W$ explicitly breaks $\IZ_{12}\to \IZ_2$ and thus there is an (anomalous) family of theories associated to the identifications 
\eq{
{\rm I.)} ~g_2\sim e^{2\pi i/3}g_2~,~~\Delta S=\frac{2\pi i}{3}\int \CP(B_2)\qquad\text{and}\qquad {\rm II.)}~ g_3\sim -g_3~,\quad \Delta S=0~ 
}
that we will discuss further below.

As discussed in~\cite{Argyres:1995jj}, the IR theory with the $\CN =1$ deformation has superpotential
\eq{W=W_{\CN =2}+\Delta W= A_1(u,v) \tildeM_1 M_1+A_2(u,v)\tildeM_2 M_2+g_2 u+g_3 v~.
}
The supersymmetric vacua of the IR theory are then given by the solutions to the $F$-term equations: $\partial _uW=\partial _vW=\partial _{M_I}W=\partial _{\tilde M_I}W=0$,
with the D-term constraint $|M_I|=|\tildeM_I|$. There are 5 solutions to these equations. Three of these are where both of the $I=1,2$ monopoles are massless and condense, $A_{I}(u,v)=0$, $M_I\tilde M_I\neq 0$, and then both $U(1)^2$ gauge fields get a mass.  These vacua have $v=0$ and are fixed under the $\IZ _4$ action; such vacua exist for any $m\equiv g_2\neq 0$, including $g_3=0$.   The $\IZ_3$ anomaly in \eqref{ADZ3anomaly} is matched by the domain walls which interpolate between the different $\IZ_2$-fixed $\CN=1$ vacua. 

For $g_3\neq 0$ there are two additional vacua with  $v\neq 0$, at points where $\partial_u A_I,\partial_v A_I=0$ for one of the $A_I$. The corresponding $M_I,\tildeM_I=0$ and the IR theory has the associated, gapless, $U(1)_I$ photon.  In the limit where $g_2\to 0$, these vacua approach the Argyres-Douglas points along $u=0$ where the theory is described by a $U(1)$ gauge theory where there are massless electrons, magnetic monopoles, and dyons. At the Argyres-Douglas points, there is an exact $\IZ_3\subset \IZ_{12}$ global symmetry that acts as electric-magnetic-dyonic triality which exchanges the three types of charged states. The theory at the Argyres-Douglas point can match the $\IZ_3$ anomaly in \eqref{ADZ3anomaly} because it is gapless \cite{Argyres:1995jj}. 

If we deform away from the $g_2=0$ and $g_3\neq 0$ Argyres-Douglas points via taking $g_2\neq0$, the phase of $g_2$ parameterizes a family of theories near the Argyres-Douglas points, explicitly breaking the $\IZ_3$ global symmetry as $g_2\sim g_2\times e^{2\pi i/3}$. 
 This gives a family anomaly due to the anomalous $\IZ_3$ global symmetry in \eqref{ADZ3anomaly}:
 \eq{
 \CA= \int d\theta\cup \CP(B_2)\quad, \quad g_2=e^{i \theta}|g_2|\quad, \quad \theta\sim \theta+\frac{2\pi }{3}~. 
}
In deforming away from the Argyres-Douglas point, the $\IZ_3$-fixed vacua are described by $\CN=1$ SUSY QED (in some duality frame). The different deformation directions correspond to either the electron, monopole, or dyonic field being the lowest mass -- which can all be brought to a frame which is the standard SQED.  The family anomaly is then matched by level crossing where the lowest mass particle changes electromagnetic charge. This level crossing is a phase transition in the IR theory, as is implied by the family anomaly. 

\vspace{0.25cm}
Starting from the $SU(3)$ theory, we can gauge the $\IZ_3^{(1)}$ center symmetry to obtain the $(PSU(3))_{k=0,1,2}$ theories, which are related by rotating  $\theta _{YM}\to \theta _{YM}+2\pi$~\cite{Aharony:2013hda}. The curve for the $\CN =2$ theory will still have three $\IZ_4$-preserving vacua at $v=0$ with two mutually-local massless monopoles/ dyons, which will condense upon deforming to $\CN =1$ by $\Delta W$.  As seen from the charges of the Wilson and 't Hooft line operators~\cite{Aharony:2013hda}, two of these $\CN=1$ vacua at $v=0$ will be trivially gapped, with the $U(1)^2$ gauge group completely broken and an unbroken $\IZ_3^{(1)}$ global symmetry, while the third vacuum has a $\IZ _3$ gauge theory and the $\IZ_3^{(1)}$ global symmetry is spontaneously broken.    These vacua are related by a non-invertible family relation as in Section \ref{sec:SU2SO3}. 

In the $PSU(3)$ theory, the anomalous family of theories near the Argyres-Douglas points are promoted to a non-invertible family parametrized by $g_2$ for fixed $g_3$. As $g_2\to 0$, the $\IZ_3$-non-invertible family relation becomes a non-invertible triality symmetry.

\subsubsection{Higher Deformations}

Now consider deforming $SU(3)$ $\CN=2$ SYM by the $\CN=1$ breaking superpotential 
\eq{
\Delta W_{UV}=\half g_4 \Tr[\Phi^4]~. 
}
This deformation is dangerously irrelevant: although irrelevant at weak coupling, in the IR it leads to non-trivial effects.  
It follows from symmetries and holomorphy \cite{Seiberg:1993vc} that $\Delta W$ is not renormalized.  Because $SU(3)$ does not have an independent quartic Casimir, in the IR the superpotential becomes
\eq{
\Delta W_{IR}=g_4 u^2~,
}
which explicitly breaks $\IZ_{12}\to \IZ_4$ and gives a mass to the IR-free chiral superfield $u$.  Since the $\IZ_3$ that is explicitly broken by $g_4\neq 0$ carries an anomaly, there is a family of theories parametrized by the phase of $g_4$, with the family anomaly: 
\eq{
\CA= \int d\theta\cup \CP(B_2)\quad, \quad g_4=e^{-i \theta}|g_4|\quad, \quad \theta\sim \theta+\frac{2\pi}{3}
}

The $F$-term equations for unbroken $\CN =1$ supersymmetric vacua are now
\eq{
&\partial_u A_1 \,\tildeM_1M_1+\partial_u A_2 \,\tildeM_2M_2+g_4 u=0~,\\
&\partial_v A_1 \,\tildeM_1M_1+\partial_v A_2 \,\tildeM_2M_2=0~,\\
&A_1M_1=0\qquad, \qquad A_2M_2=0~.
}
The three gapped $\CN =1$ vacua of the $g_2\neq 0$ deformation in the $g_4=0$ theory are also vacua for $g_4\neq 0$: the $M_I\tilde M_I$ monopole expectation values are merely appropriately adjusted to still satisfy $\partial _uW=0$ for $g_4\neq 0$.   For $g_2=0$, the $g_4\neq 0$ theory moreover has a $\CN =1$ moduli space of vacua, given by arbitrary $v$ and with $U(1)^2$ massless photons, where the $u$ field and the monopoles are massive and integrated out: 
\eq{
u=0\quad, \quad M_1=\tildeM_1=M_2=\tildeM_2=0~,
}
The IR theory on this vacuum manifold consists of $\CN =1$ supersymmetric, $U(1)^2$ gauge theory, with kinetic term coefficients $\tau _{IJ}$ given by the $\CN =2$ curve at $u=0$, as a function of the expectation value of $v$. 
This vacuum branch is the fixed locus of the $\IZ_3\subset \IZ_{12}$ action on $u,v$ and thus the coupling constant anomaly is  matched by a $\theta$-term since the UV $\IZ_3^{(1)}$ center symmetry is matched by $\IZ_3^{(1)}\subset U(1)^{(1)}\times U(1)^{(1)}$ the IR 1-form center symmetry. This vacuum manifold intersects the Argyres-Douglas points. 

\vspace{0.25cm}
Now consider further deforming $SU(3)$ $\CN=2$ SYM by the UV superpotential 
\eq{
\Delta W_{UV}=\half g_4 \Tr[\Phi^4]+\frac{1}{9} g_6\Tr[\Phi^3]^2~, 
}
breaking $\IZ_{12}\to \IZ_2$.  In the IR theory, this leads to 
\eq{
\Delta W_{IR}=g_4 u^2+g_6 v^2~, 
}
which gives a mass to $v$ in the IR, lifting the moduli space of vacua of the $g_6=0$ theory.  The $\CN =1$ supersymmetric vacuum 
equations now admit six isolated solutions.  There are the three $\IZ_2$-fixed points given by the monopole/dyon points of the $\CN=2$ theory, and the two Argyres-Douglas points that are fixed by the $\IZ_3$ action.  The sixth vacuum is a fixed point of the full $\IZ_6$ at $u,v=0$: the IR theory consists of $\CN =1$ SUSY $U(1)^2$ Maxwell theory, with coupling $\tau _{IJ}$ given by the $\CN =2$ curve at $u=v=0$.  There the coupling constant anomaly is matched by the $\theta$-term.

\section*{Acknowledgements}

We thank Jacques Distler, Dan Freed, Kantaro Ohmori, Salvatore Pace, Jaewon Song, and Ryan Thorngren for helpful comments, and especially Thomas Dumitrescu and Aiden Sheckler for helpful discussions and related collaborations.  We thank the KITP, KAIST, the Simons Center for Geometry and Physics, and the Newton Institute for hospitality in hosting workshops, and the workshop organizers and participants for stimulating discussions. 
DB and KI are supported in part by the Simons Collaboration on Global Categorical Symmetries, and by Simons Foundation award 568420.  KI is also supported by DOE award DE-SC0009919.

\appendix

\section{Constraints on Non-Invertible SPT Phases from Anomalies}
\label{app:NonInvertible}

In this appendix, we determine when a non-invertible family anomaly is trivializable by considering how a $d$-dimensional SPT phase $\IZ_N^{(p-1)}\times \IZ_N^{(p-1)}$ global symmetry for $d=2p$ transforms under a non-invertible family relation as discussed in Section \ref{sec:NonInvertibleAnomalyConstraints}.  We consider the case where the non-invertible family structure is given by 
\eq{
\CT_\theta\cong \CS \CT_{\theta+2\pi}
} 
where the $\CS$ gauging relation is given by 
\eq{
\CS Z[M_d;A_p,B_p]=\frac{1}{\sqrt{|H^p|}}\sum_{b_p\in H^p(M_d;\IZ_N)} Z[M_d;A_p,b_p]~e^{\frac{2\pi i }{2N}\int L \CP(B_p)+2 R\, B_p\cup b_p+M \CP(b_p)}~,
}
where here $\CP(B_p)$ is the Pontryagin square for $N$ even and $\CP(B_p)=B_p\cup B_p$ for $N$ odd, $L,M\in \tilde\gamma(N)\IZ_{2 N}$, and $R\in \IZ_{N}$ where 
\eq{
\gamma(N)=\begin{cases}
1&N-{\rm odd}\\
2&N-{\rm even}
\end{cases}
\quad, \quad \tilde\gamma(N)=\begin{cases}
2&N-{\rm odd}\\
1&N-{\rm even}
\end{cases}
}
We will additionally assume the theory has a non-invertible family anomaly 
\eq{
\CS Z_{\theta}[M_d;A_p,B_p]=Z_{\theta+2\pi}[M_d;A_p,B_p]~e^{-\frac{2\pi i K}{2N}\int \CP(A_p)}~,
}
where again $K\in\tilde\gamma(N) \IZ_{2N}$. Let us determine whether or not the SPT phase 
\eq{
Z_{SPT}[M_d;A_p,B_p]=e^{\frac{2\pi i }{2N}\int \left(2 k_1A_p\cup B_p+k_2\CP(B_p)\right)}~,
}
can match the non-invertible family anomaly for some coefficients $k_1\in \IZ_N$ and $k_2\in\tilde\gamma(N) \IZ_{2 N}$. 
 
 \vspace{0.25cm}
 We first consider even $N$, where $\tilde\gamma(N)=1$. We can now compute 
\eq{
&\CS Z_{SPT}[A,B]=\frac{1}{\sqrt{|H^p|}}\sum_{b\in H^p} e^{\frac{2\pi i}{2N}\int 2k_1 A\cup b+(k_2+M)\CP(b)+2R B\cup b+L\CP(B)}\\
&=\frac{1}{\sqrt{|H^p|}}\sum_{b\in H^p}e^{\frac{2\pi i (k_2+M)}{2N}\int \CP(b+\ell k_1 A+\ell R B)+\frac{2\pi i }{2N}\int (L-\ell R^2 )\CP(B)-2\ell k_1 R A\cup B-\ell k_1^2\CP(A)}\\
&= e^{\frac{2\pi i }{2N}\int (L-\ell R^2)\CP(B)-2\ell k_1 R A\cup B-\ell k_1^2 \CP(A)}
}
 where $\ell(k_2+M)=1~{\rm mod}_{2N}$. For the SPT to match the family anomaly, $k_1,k_2$ must obey
 \eq{
2 k_1=-2k_1\ell R~{\rm mod}_{2N}~,\quad L-\ell R^2=k_2~{\rm mod}_{2N}~,\quad k_1^2 \ell=K~{\rm mod}_{2N}~. 
}
We can simplify the first equation to $\ell R=-1~{\rm mod}_{N}$. This implies that both $\ell, R$ are co-prime with $N$ and therefore that they are odd numbers: $\gamma(R)=\gamma(\ell)=1$. To solve the rest of the equations it is convenient to first choose a lift of $R$ mod$_{2N}$ so that the first equation above becomes $\ell R=-1+\alpha N~{\rm mod}_{2N}$ for $\alpha=0,1$. Since $R\sim R+N$, the fact that $\ell$ is odd implies that $R\mapsto R+N$ shifts $\alpha\mapsto \alpha+1$ and we can without loss of generality set $\alpha=0$. Now we can reduce the other equations \eq{
k_2+M=-R~{\rm mod}_{2N}~,\quad L+R=k_2~{\rm mod}_{2N}\quad, \quad k_1^2=-R K~{\rm mod}_{2N}~. 
}
These have an exact solution along with a constraint:
\eq{\label{Nevenappconst}
k_2=-M-R~{\rm mod}_{2N}~,\quad L+2R+M=0~{\rm mod}_{2N}~,\quad k_1^2=-R K~{\rm mod}_{2N}~. 
}

For $N$ odd, $\tilde\gamma(N)=2$ and all of the coefficients effectively become $\IZ_N$ quantized. To facilitate this, we rewrite the SPT phase, anomaly, and gauging as
 \eq{
 \CS Z[M_d;A_p,B_p]&=\frac{1}{\sqrt{|H^p|}}\sum_{b_p\in H^p(M_d;\IZ_N)} Z[M_d;A_p,b_p]~e^{\frac{2\pi i }{N}\int L \CP(B_p)+ R\, B_p\cup b_p+M \CP(b_p)}~,\\
 Z_{SPT}[M_d;A_p,B_p]&=e^{\frac{2\pi i }{N}\int k_1 A_p\cup B_p+k_2B_p\cup B_p}~,\\
 \CS Z_{\theta}[M_d;A_p,B_p]&=Z_{\theta+2\pi}[M_d;A_p,B_p]~e^{-\frac{2\pi i K}{N}\int \CP(A_p)}~.
 }
We then obtain 
\eq{
\CS Z_{SPT}[A,B]&=\frac{1}{\sqrt{|H^p|}}\sum_{b\in H^p} e^{\frac{2\pi i}{N}\int k_1 A\cup b+(k_2+M)b^2+R B\cup b+LB^2}\\
&= e^{\frac{2\pi i }{N}\int (L-\ell^2 R^2 (k_2+M))B^2-\ell k_1 R A\cup B-\ell^2 k_1^2 (k_2+M)A^2}
}
where $x^2=x\cup x$, $2\ell(k_2+M)=1~{\rm mod}_N$, and we have dropped all gravitational terms. This implies that the SPT can match the anomaly provided $k_1,k_2$ obey the equations
\eq{
k_1=-k_1 \ell R~{\rm mod}_N~, \quad L-\ell^2 R^2 (k_2+M)=k_2~{\rm mod}_N~, \quad k_1^2 \ell^2 (k_2+M)=K~{\rm mod}_N~.
}
These can be simplified: the first equation implies $\ell R=-1~ {\rm mod}_N$ which can be used to reduce the other equations 
\eq{
L-2 k_2-M=0~ {\rm mod}_N\quad, \quad k_1^2 \ell(k_2+M)=-R K~ {\rm mod}_N~.
}
Using $2\ell (k_2+M)=1~ {\rm mod}_N$, we can also further reduce the final equation and find a second equation for $k_2$:
\eq{\label{appeq1}
k_1^2=-2 R K~{\rm mod}_2\quad, \quad 2k_2=-2M-R~ {\rm mod}_N~.
}
Using the second equation in \eqref{appeq1}, we can solve for $k_2$ and derive an additional constraint so that the full set of solutions are:
\eq{\label{finalappeqs}
k_2=-M-\frac{R-\gamma(R) N}{2}~ {\rm mod}_N~,\quad k_1^2=-2 R K ~ {\rm mod}_N~,\quad L+R+M=0~ {\rm mod}_N~.
}
These solutions are compatible with the case of $N$ even in \eqref{Nevenappconst}. This can be seen by realizing that we redefined $k_2,M,K,L\mapsto 2 k_2,2M,2K,2L$ in the even case which straightforwardly gives the first and third equations in \eqref{finalappeqs}. The second equation can also be achieved by noting that $k_1\in \IZ_N$ so that  $k_1\sim k_1+N$ which implies that shifting $k_1\mapsto k_1+N$ shifts $k_1^2\mapsto k_1^2+N$ mod$_{2N}$ for $N$ odd and therefore the equation is only meaningful mod$_N$. 

Together, the solutions in \eqref{Nevenappconst} and \eqref{finalappeqs} can be combined into the equations 
\eq{
k_2=-M-R~{\rm mod}_{2N}\quad, \quad k_1^2=-R K~{\rm mod}_{\gamma(N) N}\quad, \quad L+2R+M=0~{\rm mod}_{2N}~,
}
with the understanding that $k_2,L,M,K\in 2\IZ_{2N}$ for $N$ odd and $k_2,L,M,K\in \IZ_{2N}$ for $N$ even.   In this case, the putative non-invertible family anomaly is trivial, as it can be matched by gapped SPT counterterms.  

The non-invertible family anomaly is a genuine anomaly when these trivializing conditions do {\it not} hold. 
 
\bibliographystyle{utphys}
\bibliography{HigherGroupsCCBib}

\end{document}